\documentclass[12pt]{article}

\usepackage{amsmath,amssymb,graphicx,amsthm}
\usepackage{hyperref}
\usepackage{a4wide}
\numberwithin{equation}{section}

\begin{document}
\newcommand{\der}{\partial}
\newcommand{\e}{\mathrm{e}}
\newcommand{\ii}{\mathrm{i}}
\newcommand{\dd}{\mathrm{d}}
\newcommand{\vf}{\varphi}
\newcommand{\eps}{\varepsilon}
\newcommand{\RR}{\mathbb R}
\newcommand{\NN}{\mathbb N}
\newcommand{\mM}{\mathcal{M}}
\newcommand{\mL}{\mathcal{L}}
\newcommand{\tend}{\rightarrow}
\newcommand{\then}{\Rightarrow}
\newcommand{\z}{\hat y}
\newcommand{\Tr}{\mathrm{Tr}}
\newcommand{\sla}{\!\!\!\!\slash}
\newcommand{\ve}[2]{e^{#1}_{\phantom{#1}{#2}}}
\newcommand{\veps}[2]{{\varepsilon}^{#1}_{\phantom{#1}{#2}}}
\newcommand{\sigmahat}[2]{{\sigma}^{\hat #1}_{\phantom{#1}{\hat #2}}}
\newcommand{\sigmatilde}[2]{{\sigma}^{\tilde #1}_{\phantom{#1}{\tilde #2}}}
\newcommand{\vE}[2]{E^{#1}_{\phantom{#1}{#2}}}
\newcommand{\Sigmahat}[2]{{\Sigma}^{\hat #1}_{\phantom{#1}\hat {#2}}}
\newcommand{\Sigmatilde}[2]{{\Sigma}^{\tilde #1}_{\phantom{#1}{\tilde #2}}}
\newcommand{\sth}{\sigma_{\hat 3}}
\newcommand{\rhotilde}{\tilde\rho}
\newcommand{\be}{\begin{equation}}
\newcommand{\ee}{\end{equation}}
\newcommand{\Ypq}{Y^{p,q}}

\title{Supersymmetric asymptotically $AdS_5 \times Y^{p,q}$ solutions\\
and their CFT duals}
\date{}

\pagestyle{empty}
\begin{center}
{\Large\bf Half BPS states in $AdS_5 \times Y^{p,q}$}
\\[2.1em]

\bigskip

{\large Edi Gava$^{a,b,c}$, Giuseppe Milanesi $^{b,c,d}$,\\
 K.S. Narain$^a$ and Martin O'Loughlin$^e$}\\

\null

\noindent 
  {\it $^a$ High Energy Section,
The Abdus Salam International Centre for Theoretical Physics,
Strada Costiera 11, 34014 Trieste, Italy}
\\[2.1ex]
{\it $^b$ Istituto Nazionale di Fisica Nucleare, Sezione di Trieste}\\[2.1ex]
{\it $^c$ Scuola Internazionale Superiore di Studi
Avanzati, \\
Via Beirut 2-4, 34014 Trieste, Italy}
\\[2.1ex]
{\it $^d$  Institut f\"ur Theoretische Physik,\\
ETH Z\"urich,
CH-8093 Z\"urich, Switzerland  }
\\[2.1ex]
{\it $^e$ University of Nova Gorica,\\
Vipavska 13, 5000, Nova Gorica, Slovenia}\\
 \vfill

\end{center}

\begin{abstract}
We study a class of solutions of IIB supergravity
which are asymptotically $AdS_5\times Y^{p,q}$. They have an 
$\RR\times SO(4)\times SU(2)\times U(1)$ isometry and preserve
half of the 8 supercharges of the background geometry. 
They are described by a set of second order
differential equations that we have found 
and analysed in a previous paper, where we studied 1/8 BPS 
states in the maximally supersymmetric $AdS_5\times S^5$
background. 
These geometries correspond to certain chiral primary operators
of the $\mathcal N=1$ superconformal quiver theories,  dual
to IIB theory on  $AdS_5\times Y^{p,q}$.
 
We also show how to recover the $AdS_5\times Y^{p,q}$ backgrounds
by suitably doubling the
number of preserved supersymmetries.
We then solve the differential equations 
perturbatively in a  large $AdS_5$ radius expansion,
imposing asymptotic  $AdS_5\times Y^{p,q}$ boundary 
conditions. 
We compute the global baryonic and mesonic charges, including the
R-charge. As for the computation of the mass,
i.e. the conformal dimension $\Delta$ of the dual field
theory operators, which is 
notoriously subtle in asymptotically $AdS$ backgrounds, we
adopt the general formalism due to Wald and collaborators, which gives
a finite result, and verify the relation
$\Delta=3 R/2$,  demanded by the $\mathcal N=1$ superconformal 
algebra.  
\end{abstract}
\vfill 
\rightline{SISSA 50/2007/EP}
\newpage
\pagestyle{plain}
\section{Introduction}
One of the most impressive checks of the $AdS/CFT$ has been obtained a
few years ago \cite{bubbling}, where a very precise correspondence between
supergravity geometries and states in the dual $SU(N)$ $\mathcal N=4$ Yang-Mills theory
on $\RR\times S^3$ has been established at the 1/2 BPS level. 
More precisely, the free-fermion picture arising in the large N gauge theory
reduced on $S^3$ and restricted to the 1/2 BPS sector, has been shown to 
appear quite precisely in the exact solution of the 1/2 BPS geometries
on the supergravity side.  This goes beyond the giant graviton regime,
which corresponds to probe D3 branes wrapped on $S^3$'s either
in $AdS_5$ or $S^5$ \cite{Suryanarayana:2004ig,McGreevy:2000cw,Grisaru:2000zn,Hashimoto:2000zp} , 
in the sense that it captures the full  gravitational backreacted geometry.  
Attempts to generalise this picture to less supersymmetric geometries/states 
appeared recently in  \cite{Liu:2004ru,Chong:2004ce,Donos:2006iy,Donos:2006ms,Kim:2005ez,Chen:2007du,Gava:2006pu}. An important class of
non-local normalizable states (Wilson lines) and the corresponding dual
geometries were studied in \cite{Yamaguchi:2006te, Lunin:2006xr}

 Of course, another, but related, direction to explore would be to consider BPS states
 in less supersymmetric bulk theories. Interesting examples are
 the dual pairs given by  string theory on $AdS_5\times Y^{p,q}$  and certain $\mathcal N=1$
 Superconformal Quiver Gauge Theories, which have been subject of intense  study recently.
In \cite{Gauntlett:2004yd,Martelli:2004wu} the explicit metric on a class  of Sasaki Einstein manifolds $Y^{p,q}$ was constructed.
 A direct generalisation of the $AdS/CFT$ correspondence relates Type IIB 
String Theory on $AdS_5\times Y^{p,q}$, with $\mathcal N =1$ Quiver Gauge Theories 
\cite{Benvenuti:2004dy}. The parameters are identified as follows
\begin{equation}
	\frac{L_{AdS}^2}{4\pi \ell_s^2}=\left(\frac{\lambda}
{4\pi}\frac{\pi^3}{Vol(Y^{p,q})}\right)^{\frac 12}\qquad g_s = \frac \lambda N\,.
\end{equation}
Every $Y^{p,q}$ manifold has an $SU(2)\times U(1)\times U(1)$ isometry group 
and the $AdS_5\times Y^{p,q}$ solutions preserve 8 of the original 32 supersymmetries of type IIB supergravity. 
Supersymmetric branes wrapping cycles in $\Ypq$ have been analysed in the probe approximation in 
\cite{Arean:2004mm,Canoura:2005uz} and they may be considered as generalisations of giant gravitons.
Dual giant gravitons were studied in \cite{Martelli:2006vh,Basu:2006id}.  
A distinguishing feature of the $Y^{p,q}$ manifolds, unlike $S^5$, 
is the presence of a non-trivial 3-cycle. D3-branes can thus wrap such a non trivial 
cycle and be stable: such branes are dual to baryons in the gauge theory, the 
so called dibaryons, which are built out of products of $N$ chiral superfields 
\cite{Gubser:1998fp}. Correspondingly, on the supergravity side there is a gauge 
field coming from the four-form Ramond-Ramond gauge field, which is dual to the baryonic
current of the Gauge Theroy. 

In the quiver gauge theories associated to $Y^{p,q}$ manifolds, there are
$2p$  $SU(N)$ gauge groups and 4 types of chiral superfields,
$X$, $Y$, $U_i$ and $V_i$, $i=1,2$  in the bifundamental of $SU(N)\times SU(N)$, 
with the precise gauge assignments
encoded in the quiver diagram. The fields $U$ and $V$ are furthermore
doublets of an $SU(2)$ flavour symmetry.  
With a generic superfield $A_{\alpha}^{\beta}$,  $\alpha\in {\bf N}$
and $\beta\in \bar{\bf N}$, in the bifundamental of $SU(N)\times SU(N)$,  
one can construct dibaryonic gauge singlets
$\epsilon_{\alpha_1,\dots,\alpha_N}\epsilon^{\beta_1,\dots,\beta_N}A^{\alpha_1}_{\beta_1}\cdots A^{\alpha_N}_{\beta_N}$
The dibaryons constructed with the $SU(2)$ doublets $U_i$ and $V_i$ are furthermore in the $N+1$ dimensional 
representation of $SU(2)$. In addition to baryonic-like operators one can construct 
also mesonic-like operators, which are neutral under the baryonic charge. These are the
precise analogs of giant gravitons of the ${\mathcal N}=4$ theory. 
In any case, since our geometries preserve an $SU(2)$, in addition to $\RR\times SO(4)\times U(1)$,
they correspond to $SU(2)$ singlet operators on the gauge theory side, 
e.g. those constructed with the chiral superfields $X$ and $Y$.      
The three $U(1)$ charges, i.e. the R-charge, a flavour $U(1)$ and the baryonic charge, 
will appear as integration constants in our asymptotic solutions.

In \cite{Gava:2006pu} solutions of the type IIB equations of motion with non 
trivial R-R 5-form and $\RR\times SO(4)\times SU(2)\times U(1)$ 
isometry group preserving 4 supercharges have been studied. 
$AdS_5\times Y^{p,q}$ geometries are clearly contained in this class: 
the $\RR\times SO(4)$ is the non compact version of $U(1)\times SO(4) \subset SO(2,4)$, 
while the $SU(2)\times U(1)\times U(1)$ isometry group of $Y^{p,q}$ is contained 
in the generic $SU(2)\times U(1)$ bosonic symmetry.

In this paper we  first show in detail how to recover the $AdS_5\times Y^{p,q}$ 
geometries from the generic solutions studied in \cite{Gava:2006pu}
by requiring that additional 4 supercharges be preserved. 
We then study  1/2 BPS excitations of such geometries, namely generic 1/8 BPS solutions of
type IIB supergravity with $AdS_5\times Y^{p,q}$ asymptotics and 
$\RR\times SO(4)\times SU(2)\times U(1)$ isometry:  they represent an expansion 
of the fully backreacted geometries of D3 branes in $AdS_5\times Y^{p,q}$ . 
The brane source is substituted by flux in the same spirit as in the original \cite{bubbling}. 
Such geometries carry three net global $U(1)$ charges which are dual to the R-charge, 
a $U(1)$ flavour charge and the baryonic charge of the gauge theory. 
They are determined by four scalar functions defined on a halfspace which solve 
four nonlinear coupled differential equations. In order to specify the asymptotics and charges of the solutions we solve such equations perturbatively at large $AdS_5$ radius. The zeroth order fixes the metric and the RR 5-form as needed to describe correctly the $AdS_5\times Y^{p,q}$ geometries, the first subleading corrections determine the aforementioned global $U(1)$ charge and the second subleading correction is necessary to obtain the value of the mass. 
Solutions which carry only $R$-charge have been studied in \cite{Buchel:2006gb} at the linearised level. 

The definition of mass is somewhat subtle in asymptotically $AdS$ spacetimes,
\cite{Ashtekar:1999jx, Hollands:2005wt}
but it is even subtler when one is dealing with states in asymptotically
$AdS_5\times X^5$, with compact $X^5$, due to the fact
that the subleading terms in the metric, that in principle can be used to 
determine the mass, mix the $AdS_5$ and $M_5$  coordinates. We deal with
this problem by adopting a 10-dimensional version of the  
general construction of \cite{Wald:1999wa}, to find the conserved 
Hamiltonian and thus the correct definition of the mass. We then determine the mass
of our states and check that the BPS condition, 
relating the mass to the R-charge, is indeed satisfied by our asymptotic solutions. 

The paper is organised as follows. In Section \ref{oldstory} we give a brief summary of the results of \cite{Gava:2006pu}. In Section \ref{ads5ypq} we show how to obtain the $AdS_5\times Y^{p,q}$ geometries from the general solutions. In Section \ref{subcorr} we solve the system of differential equations up to second order in large $AdS_5$ radius (the details of the second order solutions are showed in Appendix A). In Section \ref{charges} we show how to obtain the $R$ charge and the $U(1)$ flavour charge of the solutions. 
In Section \ref{5form} we discuss subleading corrections to the RR 5-form and derive the baryon charge of the solutions. In Section \ref{asympmass} we discuss how to 
correctly define the mass for a space-time which is asymptotically a product
with an $AdS_5$ factor.  Finally, in Section \ref{concl} we present some conclusions.

\section{Description of 1/8 BPS States}\label{oldstory}
Generic solutions of type IIB Supergravity preserving 4 of the 32 supersymmetries of the theory and an $\RR\times SO(4)\times SU(2)\times U(1)$ bosonic symmetry have been constructed perturbatively in \cite{Gava:2006pu}. The metric takes the form
\begin{multline}
\dd s^2 = - h^{-2} (\dd t + V_i \dd x^i)^2 + h^2 \frac{\rho_1^2}{\rho_3^2}(T^2\delta_{ij}\dd x^i\dd x^j + \dd y^2)+\tilde\rho^2\dd\tilde\Omega_3^2+ \\
+ \rho_1^2 \big((\sigma^{\hat 1})^2 + (\sigma^{\hat 2})^2\big) +\rho_3^2(\sigma^{\hat 3} - A_t \dd t - A_i \dd x^i)^2
\label{metricBPS}
\end{multline}
with $i=1,2$; the coordinate $y$ is the product of two of the radii,
\begin{equation}
y=\rho_1 \tilde \rho > 0\,.
\end{equation}
and the function $h$ is given by
\begin{equation}
h^{-2} = \tilde{\rho}^2 + \rho_3^2(1 + A_t)^2\,.
\end{equation}
The space is a fibration of a squashed 3-sphere (on which the $SU(2)$ left-invariant 1-forms $\sigma^{\hat a}$ are defined) and a round 3-sphere $\tilde \Omega_3$ 
(on which the $SU(2)$ left-invariant 1-forms $\sigma^{\tilde a}$ are defined) 
over a four dimensional manifold.\\
The left invariant 1-forms are given by:
\begin{equation}
\begin{array}{ll}
\sigma^{\hat 1}= -\frac 12 (\cos \hat\psi\, d\hat\theta + \sin \hat\psi\, \sin\hat \theta\, d\hat \phi) & \sigma^{\tilde 1}= -\frac12(\cos \tilde\psi\, d\tilde\theta + \sin \tilde\psi\, \sin \tilde\theta\, d\tilde\phi)\\
\sigma^{\hat 2}= -\frac12(-\sin\hat \psi\, d\hat\theta + \cos\hat \psi\, \sin\hat \theta\, d\hat \phi )& \sigma^{\tilde 2}= -\frac12(-\sin \tilde\psi\, d\tilde\theta + \cos \tilde\psi\, \sin\tilde \theta\, d\tilde\phi) \\
\sigma^{\hat 3}= -\frac12(d\hat\psi + \cos \hat\theta\, d \hat\phi) & \sigma^{\tilde 3}= -\frac12(d\tilde\psi + \cos \tilde\theta\, d \tilde\phi) 
\label{sigmas}
\end{array}
\end{equation}
%
and satisfy the relations (with $\sigma^a$ being either $\sigma^{\hat a}$ or $\sigma^{\tilde a}$)
\begin{equation}
\dd \sigma^{a} = \epsilon_{abc} \sigma^{b}\wedge \sigma^{c}\,.
\end{equation}
With this normalisation the metric on the unit radius round three sphere is given by
\begin{equation}
	d \Omega_3^2 = (\sigma^1)^2+(\sigma^2)^2+(\sigma^3)^2\,.
\end{equation}

The only non trivial field strength in our Ansatz is the Ramond-Ramond 5-form: 
it is more conveniently expressed in terms of the ``d-bein''
\begin{align}
e^0 =& h^{-1}(\dd t + V_i \dd x^i)\\
e^j =& h\frac{\rho_1}{\rho_3}T\delta^j_i\dd x^i\\
e^3=& h \frac{\rho_1}{\rho_3}\dd y\\
e^{\hat a} =& \begin{cases}
	\rho_1 \sigma^{\hat a} & \hat{a}=1,2\\
	\rho_3(\sigma^{\hat 3} -A_\mu\dd x^\mu) & \hat{a}=3
      \end{cases}\\
e^{\tilde a} =& \tilde \rho \sigma^{\tilde{a}}
\end{align}
as
\begin{multline}
	F_{(5)} =  2 \left( \tilde G_{mn} e^m\wedge e^n + \tilde V_m e^m\wedge e^{\hat 3} + \tilde g e^{\hat 1}\wedge e^{\hat 2} \right)\wedge \tilde\rho^3\dd \tilde \Omega_3+\\
	2\left(- G_{pq}e^p\wedge e^q \wedge e^{\hat 1}\wedge e^{\hat 2}\wedge  e^{\hat 3} +\star_4 \tilde V\wedge  e^{\hat 1}\wedge e^{\hat 2}- \star_4 \tilde g \wedge e^{\hat 3}\right) \,,
\end{multline}
where
\begin{gather}
	G_{mn} = \frac12 \epsilon_{mnpq} \tilde G^{pq}\\
	\star_4 \tilde V = \frac{1}{3!}\epsilon_{mnpq}\tilde V^m e^n\wedge e^p\wedge e^q\\
	\star_4 \tilde g  = \tilde g e^0\wedge e^1\wedge e^2\wedge e^3\,.
\end{gather}
The complete solution can be expressed in terms of four independent functions $m,n,p,T$ defined on the halfspace $(x^1,x^2,y)$, as follows
\begin{equation}	
\begin{array}{lll}
 \rho_1^4 =  \frac{m p+ n^2}{m} y^4& \rho_3^4 = \frac{p^2}{m(mp+n^2)}&\tilde\rho^4 = \frac{m}{mp+n^2}\\ 
  h^{4} = \frac{ m p^2}{mp+n^2}& A_t = \frac{n-p}{p} & A_i = A_t V_i -\frac12 \epsilon_{ij}\der_j \ln T
\end{array}
\end{equation}
and
\begin{gather}
	\dd V = -y \star_3 [ \dd n + (n D +2y m (n-p) + 2n/y)\dd y]\label{firstdV}\\
	\der_y \ln T = D \label{firstdT}\\
	 D\equiv 2y(m+n-1/y^2)\,,
\end{gather}
where $\star_3$ indicates the Hodge dual in the three dimensional diagonal metric
\begin{equation}
	\dd s_3^2 = T^2\delta_{ij}\dd x^i\dd x^j + \dd y^2\,.
\end{equation}

The various four-dimensional forms from which the 5-form field strength is constructed are given by
\begin{gather}
\tilde g= \frac{1}{4 \tilde\rho}\left[1-\frac{\rho_3^2}{\rho_1^2}(1+A_t)\right]\\
	\tilde V = \frac 12 \frac{1}{\rho_3 \tilde\rho^3}\dd (\tilde{g} \rho_1^2 \tilde\rho^3) \\
G \rho_1^2 \rho_3 = \dd B_t \wedge (\dd t + V_i\dd x^i) + B_t \dd V + \dd \hat B\\
\tilde G \tilde \rho^3 = \frac12 g \rho_1^2 \tilde\rho^3  \dd A + \dd \tilde B_t \wedge (\dd t + V_i\dd x^i)+\tilde B_t \dd V + \dd \hat{\tilde B}\,,
\end{gather}
with
\begin{equation}
\begin{split}
 & \tilde B_t = -\frac{1}{16} y^2\, \frac{n-1/y^2}{p}\\
& \dd \hat{\tilde B} = -\frac1{16} y^3\star_3 [\dd m + 2 m D\,\dd y]\\
& B_t = -\frac 1 {16} y^2\,\frac{n}{m}\\
& \dd \hat B = \frac 1 {16}y^3\star_3  [\dd p + 4yn(p-n)\dd y]\,.\\
\end{split}
\end{equation}
The Bianchi identities on $F_{(5)}$ and the integrability condition for \eqref{firstdV} give three second order differential equations on $m,n,p$ which, together with \eqref{firstdT} give a system of nonlinear coupled elliptic differential equations
\begin{equation}
\label{equations}
\begin{split}
&y^3 (\der_1^2+\der_2^2) n + \der_y \left(y^3 T^2 \der_y n\right) + y^2\der_y\big[ T^2 \big(y D n +2y^2m(n-p)\big)\big]+4 y^2 D T^2 n=0\\
&y^3 (\der_1^2+\der_2^2) m + \der_y \left(y^3 T^2 \der_y m\right) + \der_y \left( y^3 T^2 2m D\right)=0\\
&y^3 (\der_1^2+\der_2^2) p + \der_y \left(y^3 T^2 \der_y p\right) + \der_y \big[ y^3 T^2 4n y(n-p)\big]=0\\
&\der_y \ln T = D\,.
\end{split}
\end{equation}

\section{$AdS_5\times Y^{p,q}$ solutions}\label{ads5ypq}
Taking any solution described in Section \ref{oldstory} and assuming rotational symmetry in the $\{x^1,x^2\}$ plane, the bosonic symmetry is enhanced to $\RR\times SO(4)\times SU(2)\times U(1) \times U(1)$. We will first consider a subset of solutions which preserve 8 supersymmetries (the generic solution preserves only 4 of them as explained in the previous section). The well known $AdS_5\times Y^{p,q}$ \cite{Martelli:2004wu} are clearly contained in this subset: the round $S^3$ is a factor in $AdS_5$, as suggested by the analysis in \cite{Gava:2006pu}, with $\RR\times SO(4)$ the non compact version of $U(1)\times SO(4) \subset SO(2,4)$, while the remaining $SU(2)\times U(1)\times U(1)$ is the isometry group of the generic $Y^{p,q}$ metric.
\subsection{Constraints for enhanced supersymmetry}
Since the solutions described in \cite {Gava:2006pu} generically preserve only 4 supersymmetries, the $AdS_5\times Y^{p,q}$ geometries will be specified by a set of constraints on the four functions $m,n,p$ and $T$. We will now show how these constraints arise.

The supersymmetry parameters that leave invariant our background are the solutions to the Killing spinor equation
\begin{equation}\label{gravitino}
	\delta \chi_M = \nabla_M \psi + \frac{\ii}{480}F_{M_1 M_2 M_3 M_4 M_5} \Gamma^{M_1 M_2 M_3 M_4 M_5}\Gamma_M \psi = 0 \,.
\end{equation}
As a consequence of the symmetry assumptions we look for a solution $\psi$ of the form
\begin{equation}
	\psi = \eps \otimes \hat \chi \otimes \tilde \chi_{(b)}\,.
\end{equation}
Here $\eps$ is an 8 component complex spinor and $\hat \chi, \tilde\chi_{(b)}$ are 2 component complex spinors defined on the two 3-spheres satisfying
\begin{gather}\label{conditions}
	\frac{\der}{\der \omega^{\hat a}} \hat\chi = 0 \qquad \sigma_{\hat 3} \hat\chi = s \hat\chi\\
	\nabla'_{\tilde a}\tilde\chi_{(b)} = b \frac{\ii}{2}\sigma_{\tilde a} \tilde \chi_{(b)}
\end{gather}
where $\omega^{\hat a},\omega^{\tilde a}$ are coordinates on the two spheres. $\nabla'$ is the covariant derivative on the unit radius three sphere and $s,b=\pm 1$. The spinor $\hat \chi$ is a singlet under the $SU(2)_L$ isometry of the squashed 3-sphere, while the spinors $\tilde \chi_\pm$ transform as the $(0,\frac 12)$ for upper sign and $(\frac 12, 0)$ for the lower sign, of the $SO(4)$
isometry of the round $S^3$, which is part of $AdS_5$. 

The analysis in \cite{Gava:2006pu} fixes $b=s=1$, i.e. $\tilde\chi$
has definite chirality in $SO(4)$ and $\hat\chi$ is highest weight
of the broken $SU(2)_R$. $\eps$ is proportional to some $\eps_0$ obeying
$\eps_0^\dagger\eps_0=1$. Since we have a doublet of $\tilde\chi_{(1)}$, the space of solutions is 2 dimensional and complex giving rise to 4 real preserved supersymmetries. 
We will show that $AdS_5\times Y^{p,q}$ geometries are obtained by requiring that spinors with $b=-1$, $s=1$ are also solutions of the equations \eqref{gravitino}. In this case there are two doublets of $\tilde \chi$ and thus 8 real solutions to \eqref{gravitino}.
This agrees with what one expects from the $\mathcal N=1$ SCFT side: there, out of the 4 pairs of Killing 
spinors $\xi_{\pm}^A$, $A=1,\dots,4$ in the $\bf 4$ of $SU(4)$ of the ${\mathcal N}=4$ theory 
on $\RR\times S^3$, obeying 

\begin{equation}
D_\mu\xi_{\pm}^A=\pm \frac{i}{2}\sigma_\mu\xi_{\pm}^A\,
\label{krs3}
\end{equation}
only the $SU(3)$ singlet $\xi_\pm $  in $SU(3)\times U(1)\subset SU(4)$  survives in the $\mathcal N=1$ case. This has $SU(4)$ weights  $(\frac{1}{2},  \frac{1}{2},  \frac{1}{2})$
and, picking the $SO(4)$ inside $SU(4)$ corresponding for example  to the first
two entries, we see that it is a singlet of, say, $SU(2)_L$ and highest weight
of $SU(2)_R$ in the $SO(4)\subset SU(4)$.  Furthermore, the two  signs in (\ref{krs3})  correspond to the two chiralities  
 of the $SO(4)$ isometry group of $S^3$. Since this $S^3$ corresponds to the $S^3$ inside $AdS_5$,
 this checks with the above requirement of  $b=\pm 1$.

Due to the conditions on the spinor, $\hat \chi$ and $\tilde \chi_{(b)}$ factorise in each component of the gravitino variation equation which then becomes equivalent to the following system of coupled differential and algebraic equations on $\eps$\footnote{For example the first equation is obtained as follows
\begin{multline}
	\left(\nabla_\mu+M \Gamma_\mu \right)\psi = \left(\tilde\nabla_\mu -\frac14\rho_3 F_{\mu\nu}\Xi^\nu_{\phantom \nu m}\Gamma^m\Gamma^{\hat 3}+A_\mu\left(\Sigma_{\hat 3}+\Gamma^{\hat 1}\Gamma^{\hat 2}\right)-
A_\mu \nabla_{\hat 3}+M \Gamma_\mu\right)\psi=\\
=\left(\tilde\nabla_\mu -\frac14\rho_3 F_{\mu\nu}\Xi^\nu_{\phantom \nu m}\Gamma^m\Gamma^{\hat 3}+A_\mu\Gamma^{\hat 1}\Gamma^{\hat 2}+M\left(\Gamma_\mu+A_\mu\rho_3\Gamma_{\hat 3}\right) \right)\psi= \\
=\left(\tilde\nabla_\mu -\frac14\rho_3 F_{\mu\nu}\Xi^\nu_{\phantom \nu m}\gamma^m\sigma^{\hat 3}+A_\mu\sigma_{\hat 3}+M\gamma_\mu\right)\psi
\end{multline}}
\begin{align}
	&\left[\tilde \nabla_\mu - \frac 14 F_{\mu\nu}\Xi^\nu_{\phantom \nu m} \gamma^m\gamma^5 \hat \sigma_1 s+ \ii A_\mu s - \bigg( \tilde G \sla + \tilde V \sla \gamma_5 \hat \sigma_1 s + \ii \tilde g s\bigg) \gamma_5\hat \sigma_2 \gamma_\mu \right]\eps=0\label{mu}\\
&\left[\frac \ii 2 \frac{\rho_3}{\rho_1}\gamma_5\hat\sigma_1+\frac12 \der\,\sla\rho_1+\rho_1\bigg( \tilde G \sla + \tilde V \sla \gamma_5 \hat \sigma_1 s  - \ii \tilde g s \bigg) \gamma_5\hat \sigma_2\right]\eps=0\label{a12}\\
&\left[\frac{\ii}{2}\left(2-\frac{\rho_3^2}{\rho_1^2}\right) \gamma_5\hat \sigma_1 +\frac12 \der\,\sla \rho_3+\frac18 \rho_3^2 F\sla\gamma_5\hat\sigma_1 s +\rho_3\bigg( \tilde G \sla - \tilde V \sla \gamma_5 \hat \sigma_1 s  + \ii \tilde g s \bigg) \gamma_5\hat \sigma_2 \right]\eps=0\label{a3}\\
&\left[\frac \ii 2 b\gamma_5\hat\sigma_2 + \frac12 \der\,\sla \tilde \rho -\tilde\rho \bigg( \tilde G \sla + \tilde V \sla \gamma_5 \hat \sigma_1 s  + \ii \tilde g s \bigg)\gamma_5\hat\sigma_2\right]\eps=0\label{at}. 
\end{align}
Note that the first equation is a  first order differential 4-vector equation for $\eps$ while the last three are  algebraic 4-scalar equations.

We now express all the supergravity fields via the functions $m,n,p$ and $T$ of the previous section. We are thus guaranteed that a solution to the above system with $b=s=1$ exists by the analysis in \cite{Gava:2006pu}. We now ask that a second solution to these equations exists for $b=-1$, $s=1$.

We have used Mathematica to solve explicitly the equations. The existence of solutions
implies certain constraints on the background, which are more conveniently expressed in terms of the metric entries as
\begin{equation}\label{constraint}
\begin{split}
	&1+A_t = \rho_1^2/\rho_3^2 \\
	&\rho_1^2 - \rho_3^2 = S^4/\rho_1^2\\
	&T^2 \der_y \ln (\rho_1/\rho_3)\big[2 \rho_1^2/\rho_3^2 -2 + y \der_y \ln(\rho_1/\rho_3)\big]+y \big[\der_r \ln(\rho_1/\rho_3)\big]^2=0
\end{split}
\end{equation}
where $(r,\phi)$ are polar coordinates in the $(x^1,x^2)$ plane. Notice that the first two constraints together with the relation $y=\rho_1 \tilde \rho$ allow us to express the four functions $\rho_1,\rho_3,\tilde \rho, A_t$ in terms of only one function. The last constraint together with the equation for $\der_y T$ can be used to eliminate $T$.
Moreover, the three second order differential equations that came from the integrability condition for the $1/8$ supersymmetric geometries are reduced to a single equation which is more easily expressed in terms of the function $\tilde z$
\begin{gather}\label{ztilde}
	\tilde z \equiv \frac12 \left[ 1+ \tanh \left( \frac {\rho_3 (1+A_t)}{\tilde\rho}\right)\right]\\
\label{eqnztilde}\frac 1 r \der_r \big( r \der r \tilde z \big)+y \der_y \left\{ T^2 \frac 1 y \left[\der_y \tilde z + 4 \tilde z (1-\tilde z) \frac{\rho_1^2/\rho_3^2-1}{y}\right]\right\}=0
\end{gather}
where the combination $\rho_1^2/\rho_3^2$ is given by
\begin{equation}
	\frac {\rho_1^2}{\rho_3^2} = \left(\frac {1+\sqrt{1-4 S^4 \frac{1-\tilde z}{y^2 \tilde z} }}{2}\right)^{-1}
\end{equation}
and $T$ can be found by solving the third equation in \eqref{constraint}. The solution is thus specified completely by a single function
\footnote{If, instead of doubling supersymmetry by requiring $b=-1$ and $s=1$ in addition 
to $b=s=1$,  one requires
to have solutions of \eqref{gravitino} also for $b=s=-1$, then one obtains a different set of constraints on the
background. By making an asymptotic analysis similar to the one we will perform
here in Section \ref{subcorr}, it can be shown that the resulting geometry describes 1/4-BPS states in the background
$AdS_5\times S^5$}.

\subsection{$AdS_5\times Y^{p,q}$ metrics}
In this section, we are going to show how the $AdS_5\times Y^{p,q}$ geometries arise from the generic description given above. As a first step we present the conditions that should be satisfied by the $1/4$ supersymmetric solutions in order that they factorise into,
\begin{equation}
 	AdS_5\times X^5\,,
\end{equation}
For some supersymmetric five-manifold $X^5$. These will turn out to be equivalent to a single first order differential equation which implies the second order equation in \eqref{eqnztilde}. The opposite in general cannot be proven: the generic solution preserving 8 supersymmetries apparently is not factorisable in general.

As a second step we will prove, by giving the explicit coordinate transformation to the gauge in \cite{Martelli:2004wu}, that the $X^5$ factor is indeed a generic $Y^{p,q}$ manifold. 

First of all we switch to the more convenient coordinates $(\tilde \rho, \rho_1,\tilde \phi,\hat\psi')$ defined by
\begin{equation}
\begin{split}
	& y = \rho_1 \tilde \rho	\\
	& r = r (\rho_1,\tilde \rho) \\
	& \phi = \tilde \phi + \tilde c \, t \quad\\
	&\hat \psi = \hat \psi' -2 \gamma \,t\,-2 \delta \, \phi \equiv \hat\psi'-(2\gamma + 2 \tilde c \, \delta) t - 2\delta \, \tilde \phi
\end{split}
\end{equation}
Using the constraints in \eqref{constraint} the solution is completely specified once the explicit form of the function $r (\rho_1,\tilde \rho)$ is known.

The last shift implies that the left invariant one-form $\sigma^{\hat 3}$ is shifted to $\sigma^{\hat 3}\,'+(\gamma + \tilde c \,\delta)\dd t + \delta\,\dd \tilde\phi$. With a slight abuse of notation we will keep calling this shifted one form $\sigma^{\hat 3}$.
The metric of \ref{metricBPS} is thus
\begin{multline}
\dd s^2 = - h^{-2} (\dd t^2 + V_\phi \dd\phi)^2 + h^2 \frac{\rho_1^2}{\rho_3^2}( T^2 \delta_{ij} \dd x^i \dd x^j + \dd y^2) + \\
+ \tilde \rho^2 \dd \Omega_3^2 + \rho_1^2 \big[\big(\sigma^{\hat 1}\big)^2+\big(\sigma^{\hat 2}\big)^2\big]+ \rho_3^2 (\sigma^{\hat 3}-A_t\dd t - A_\phi \dd \phi)^2=\\
=g_{tt}\dd t^2 + g_{\tilde\rho\tilde\rho} \dd \tilde\rho^2 + \tilde \rho^2 \dd \tilde\Omega_3^2 
+2 g_{t\tilde \phi} \dd t \dd \tilde\phi+ 2g_{\rho_1 \tilde \rho} \dd \rho_1 \dd \tilde \rho+\\
 + g_{\rho_1 \rho_1} \dd \rho_1^2+ g_{\tilde \phi \tilde \phi} \dd \tilde \phi^2 + \rho_1^2 \big[ \big(\sigma^{\hat 1}\big)^2+\big(\sigma^{\hat 2}\big)^2\big]+\\
+ \rho_3^2 \bigg[\sigma^{\hat 3}+\big(\gamma -A_t- \tilde c(A_\phi-\delta)\big)\dd t - (A_\phi-\delta) \dd \tilde\phi\bigg]^2\,,
\end{multline}
with
\begin{equation}
\begin{split}
	&g_{tt} = - h^{-2} (1+\tilde c V_\phi)^2 +  \tilde c^2 h^2  \frac {\rho_1^2}{\rho_3^2}T^2r^2  \\
	&g_{\tilde \rho \tilde \rho} = h^2 \frac{\rho_1^2}{\rho_3^2}\left[\rho_1^2 + T^2r^2  \left(\frac{\der \ln r}{\der \tilde \rho}\right)^2\right]\\
	&g_{t\tilde\phi} = - h^{-2} (1+ \tilde c V_\phi)V_{\phi}+ \tilde c h^2 \frac{\rho_1^2}{\rho_3^2}T^2r^2  \\
	&g_{\rho_1 \tilde \rho} = h^2 \frac{\rho_1^2}{\rho_3^2} \left[\rho_1 \tilde \rho + T^2 r^2 \frac{\der\ln r}{\der\tilde\rho}\frac{\der \ln r}{\der \rho_1}\right]\\
	& g_{\rho_1 \rho_1} = h^2 \frac{\rho_1^2}{\rho_3^2}\left  [\tilde \rho^2 +  T^2r^2 \left(\frac{\der \ln r}{\der \rho_1}\right)^2\right]\\
	& g_{\tilde \phi \tilde \phi} = - h^{-2} V_\phi^2 + h^2 \frac{\rho_1^2}{\rho_3^2} T^2 r^2
\end{split}
\end{equation}
We recall the constraint on the metric components coming from the requirement of 
$1/4$ supersymmetry,
\begin{equation}
\begin{split}
	&1+A_t = \frac{\rho_1^2}{\rho_3^2}\\
	&\rho_1^2 - \rho_3^2 = \frac{S^4}{\rho_1^2}\\
	& h^{-2} = \tilde\rho^2 + \rho_1^4 / \rho_3^2.
\end{split}
\end{equation}
In order that the metric factorises we need the $\dd t \, \sigma^{\hat 3}$ term to vanish which requires that 
\begin{equation}
	A_t + \tilde c (A_\phi-\delta) = \gamma\,.
\end{equation}
Imposing also $g_{t\tilde \phi} = 0$ we obtain 
\begin{equation}\label{gttildephi}
	\tilde c h^2 \frac{\rho_1^2}{\rho_3^2} T^2 r^2 = h^{-2}( 1+ \tilde c V_\phi) V_\phi\,.
\end{equation}
In order to have an $AdS_5$ factor we should have $- g_{tt} = L^2 + \tilde \rho^2$ which gives, using the last relation
\begin{equation}
	h^{-2} (1 + \tilde c V_\phi) = L^2 + \tilde \rho^2\,,
\end{equation}
We also demand that $g_{\tilde \rho \tilde \rho} = \frac{L^2}{L^2 + \tilde \rho^2}$ which after a little bit of algebra gives
\begin{equation}\label{dlogrdrhotilde}
	\frac{\der\ln r}{\der\tilde\rho} = \pm \frac{ \tilde c \tilde \rho}{L^2 +\tilde \rho^2}.
\end{equation}
Requiring that we have a product metric means that we also must impose that $g_{\rho_1 \tilde \rho} = 0$ which implies
\begin{equation}\label{dlogrrho1}
	\frac{\der\ln r}{\der \rho_1} = \mp \tilde c \frac{\rho_1^3}{\rho_3^2 L^2 - \rho_1^4}.
\end{equation}
As a result we find immediately that 
\begin{equation}
	g_{\rho_1 \rho_1} = \frac{\rho_1^2 L^2}{\rho_3^2 L^2 - \rho_1^4}
\end{equation}
and
\begin{equation}
	g_{\tilde \phi \tilde \phi} = \frac{1}{\tilde c^2}\left(L^2 - \frac{\rho_1^4}{\rho_3^2}\right)\,.
\end{equation}

The generic solutions to equation \eqref{dlogrdrhotilde} for the upper sign are
\begin{equation}
 	r = (L^2 + \tilde \rho^2)^{\tilde c /2} r_0(\rho_1) \rho_1^{\tilde c}
\end{equation}
where we have extracted the $\rho_1^{\tilde c}$ for future convenience. \eqref{dlogrrho1} is an equation for $r_0(\rho_1)$
\begin{equation}\label{r0}
 	r_0'(\rho_1) = \tilde c \frac {L^2(\rho_1^4-S^4)}{\rho_1^7-L^2 \rho_1 (\rho_1^4-S^4)} r_0 (\rho_1)
\end{equation}
Using the last constraint in \eqref{constraint} 
\begin{equation}
 	T^2 \der_y \ln (\rho_1/\rho_3)\big[2 \rho_1^2/\rho_3^2 -2 + y \der_y \ln(\rho_1/\rho_3)\big]+y \big[\der_r \ln(\rho_1/\rho_3)\big]^2=0\,
\end{equation}
we can find $T$. Note that both the first order differential equation for $T$
\begin{equation}
 	\der_y \ln T = D
\end{equation}
and the second order equation in \eqref{eqnztilde} are satisfied when $r_0(\rho_1)$ satisfies the equation \eqref{r0}.
\subsection{Relation to standard $Y^{p,q}$ coordinates}
Now we show the coordinates transformation that brings the metric on $X_5$ to the standard metric on $Y^{p,q}$ as presented in \cite{Martelli:2004wu}.
We perform the rescaling
\begin{equation}
\rhotilde\tend L\rhotilde,\quad\quad\rho_i\tend L \rho_i,\quad\quad S\tend L S	
\end{equation}
which takes the metric of $AdS_5$ into the form
\begin{equation}
 	\dd s^2_{AdS_5} = L^2 \bigg( -(\rhotilde^2 + 1) \dd t^2 + \frac{\dd \rhotilde^2}{\rhotilde^2+1}+\rhotilde^2\dd\Omega_3^2\bigg)
\end{equation}
while the metric on the ``internal'' part is
\begin{multline}
	\dd s_5^2 = L^2 \bigg[\frac{\rho_1^2 }{\rho_3^2 - \rho_1^4} \dd \rho_1^2 + \frac{1}{\tilde c^2}\left(1 - \frac{\rho_1^4}{\rho_3^2}\right) \dd \tilde \phi^2+ \\
+ \rho_1^2 \big[ \big(\sigma^{\hat 1}\big)^2+\big(\sigma^{\hat 2}\big)^2\big]+ \rho_3^2 \big(\sigma^{\hat 3} - (A_\phi-\delta) \dd \tilde\phi\big)^2 \bigg]\,.
\end{multline}

The standard form for the metric on $Y^{p,q}$ \cite{Martelli:2004wu} is, 
\begin{equation}\label{ypqmetric}
\begin{aligned}
   \dd s^2 &= \frac{1-c\hat y}{6}(\dd\hat\theta^2+\sin^2\hat\theta
      \dd\hat\phi^2)+\frac{1}{w(\hat y)q(\hat y)}
      d\hat y^2+\frac{q(\hat y)}{9}(\dd\hat \psi+\cos\hat\theta \dd\hat\phi)^2 \\ & \qquad \qquad
      + {w(\hat y)}\left[\dd\alpha -\frac{ac-2\hat y+\hat y^2c}{6(a-\hat y^2)}
         (\dd\hat\psi+\cos\hat\theta \dd\hat\phi)\right]^2
\end{aligned}
\end{equation}
with\footnote{Notice that the $\hat\psi$ of \cite{Martelli:2004wu} has the opposite sign to that used here.}
\begin{equation}
\begin{split}
&w(\hat y) = \frac{2(a-\hat y^2)}{1-c\hat y}\\
 &q(\hat y) = \frac{a-3\hat y^2+2c\hat y^3}{a-\hat y^2}~.
\end{split}
\end{equation}

Recalling that
\begin{equation}
\begin{array}{l}
\sigma^{\hat 1}= -\frac 12 (\cos \hat\psi\, d\hat\theta + \sin \hat\psi\, \sin\hat \theta\, d\hat\phi) \\
\sigma^{\hat 2}= -\frac12(-\sin \hat\psi\, d\hat\theta + \cos\hat \psi\, \sin \hat\theta\, d\hat\phi ) \\
\sigma^{\hat 3}= -\frac12(d\hat\psi + \cos \hat\theta\, d\hat \phi) 
\end{array}
\end{equation}
we immediately get
\begin{equation}\label{rho12(y)}
	\rho_1^2 = \frac 23 (1- c \hat y)
\end{equation}
and
\begin{equation}\label{rho32(y)}
 \frac{1}{4} \rho_3^2 = \frac{2+a c^2-6 c \hat y + 3 c^2 \hat y^2}{18(1-c \hat y)}\,.
\end{equation}
Recalling that $\rho_1^2 - \rho_3^2 = S^4/\rho_1^2$ we have
\begin{equation}\label{S}
	S^4 = \frac 4 {27} (1-a c^2)\,.
\end{equation}
We also have 
\begin{equation}
	A_\phi = \frac 1 {\tilde c} ( \gamma - A_t) +\delta = \frac 1 {\tilde c} \left( \gamma +1 -\frac{\rho_1^2}{\rho_3^2}\right) + \delta\,.
\end{equation}
Assuming that $\alpha = \beta \tilde \phi$ and equating the $\dd \tilde \phi^2$ component of the metric we get
\begin{equation}
	\frac 1 {\tilde c^2} \left(1 - \frac{\rho_1^4}{\rho_3^2}\right) + \rho_3^2 (A_\phi-\delta)^2 = w(\hat y) \beta^2
\end{equation}
which implies after some straightforward algebra that
\begin{equation}
	\gamma = \frac 12\quad,\quad \beta =\pm \frac c {2 \tilde c}\,.
\end{equation}
The coefficient of the cross term $\dd \tilde \phi\, \sigma^{\hat 3}$ is
\begin{equation}
	\frac 12 \rho_3^2 (A_\phi-\delta) = - \beta w(\hat y) \frac{ac-2\hat y+\hat y^2c}{6(a-\hat y^2)}
\end{equation}
which implies that we must have $\beta = - \frac c { 2\tilde c}$. We can therefore set 
\begin{equation}
  \beta = - 1 \qquad \tilde c = \frac 12 c\,.
\end{equation}
Using the expression for $r$
\begin{equation}
\begin{split}
	 &r = (L^2 + \tilde \rho^2)^{\tilde c /2} r_0(\rho_1) \rho_1^{\tilde c}\\
	&r_0'(\rho_1) = \tilde c \frac {L^2(\rho_1^4-S^4)}{\rho_1^7-L^2 \rho_1 (\rho_1^4-S^4)} r_0 (\rho_1)
\end{split}
\end{equation}
and the definition
\begin{equation}
 	A_\phi = A_t V_\phi + \frac 12 r \der_r \ln T
\end{equation}
we get
\begin{equation}\label{At}
 	A_t+\tilde c (A_\phi - \delta) = \frac 1 2  - \frac {\tilde c} 2 (1+2 \delta)
\end{equation}
which gives
\begin{equation}
 	\delta = -\frac 12
\end{equation}
for $c,\tilde c\neq 0$.
 Finally, the matching of the $\dd \hat y^2$ factor
\begin{equation}
	\frac 1 9 \frac{c^2}{\rho_3^2-\rho_1^4} = \frac{1}{w(\hat y) q(\hat y)}
\end{equation}
is identically satisfied. Finally, we observe that the polynomial $q(\hat y)=a-3\hat y^2+2c\hat y^3$, whose zeroes
$\hat y_1$ and $\hat y_2$, with $\hat y_1< 0$ and $\hat y_2$ the smallest between the two other positive zeroes, 
determine the range of $\hat y$, $\hat y_1\leq \hat y \leq \hat y_2$, can be expressed in terms of $\rho_1$
as  $q=-27(\rho_1^6-\rho_1^4+S^4)/4$.
Notice also that in the metric \eqref{ypqmetric} any non zero value of $c$ can be reabsorbed in a rescaling of $\hat y$ and $\alpha$. 
We may thus set $c=1$ whenever $c\neq0$.

\subsection*{c=0 case}
Let us now take a look at the singular case
\begin{equation}
	c =\tilde c= 0
\end{equation}
which corresponds to the Sasaki-Einstein internal manifold $Y^{1,0}\equiv T^{1,1}$.
From the equations \eqref{gttildephi},\eqref{At} we can immediately obtain
\begin{equation}
\begin{split}
 &A_t = \gamma = \frac 12\\
&V_\phi=0
\end{split}
\end{equation}
and from \eqref{rho12(y)},\eqref{rho32(y)},\eqref{S}
\begin{equation}
\begin{split}	
&\rho_1^2 = \rho_3 = \frac 2 3\\
& S = \frac 4 {27}\,.
\end{split}
\end{equation}
Notice that the equation for $\rho_1$ implies
\begin{equation}
 	y = \sqrt{\frac 23} L^2 \rhotilde
\end{equation}
Given these explicit values for $\rho_1$ and $\rho_3$, the last constraint in \eqref{constraint} 
\begin{equation}
 	T^2 \der_y \ln (\rho_1/\rho_3)\big[2 \rho_1^2/\rho_3^2 -2 + y \der_y \ln(\rho_1/\rho_3)\big]+y \big[\der_r \ln(\rho_1/\rho_3)\big]^2=0
\end{equation}
is automatically satisfied.

%
The metric on the internal manifold becomes
\begin{equation}
	\dd s^2_5 = L^2 \left[\frac 3 2{\tau(r)^2}(\frac{\dd  r^2}{r^2} + \dd \tilde\phi^2)+ \frac 2 3   \big[ \big(\sigma^{\hat 1}\big)^2+\big(\sigma^{\hat 2}\big)^2\big]+ \frac 4 9 \big(\sigma^{\hat 3} -(A_\phi-\delta) \dd \tilde\phi\big)^2\right]
\end{equation}
where
\begin{equation}
	T^2 = (\rhotilde^2 + 1 )\frac{\tau(r)^2}{r^2}
\end{equation}
is such that $T$ solves the equation

\begin{equation}
 	\der_y \ln T = D \iff \der_{\rhotilde} \ln T = \frac {2 \rhotilde}{\rhotilde^2 +1}
\end{equation}

We now match this expression with the one in \cite{Martelli:2004wu}. For $c=0$, $a$ can be reabsorbed in a coordinate redefinition. We set, for convenience, 
\begin{equation}
 	a=3
\end{equation}
and obtain,
\begin{equation}
\begin{aligned}
   \dd s^2 &= \frac{1}{6}(\dd\hat\theta^2+\sin^2\hat\theta
      \dd\hat\phi^2)+\frac{1}{6(1-\hat y^2)}
      \dd \hat y^2+\frac{1- \hat y^2}{3(3-\hat y^2)}(\dd\hat\psi+\cos\hat\theta \dd\hat\phi)^2 \\ & \qquad \qquad
      + {2(3-\hat y^2)}\left[\dd\alpha + \frac{2\hat y}{6(3-\hat y^2)}
         (\dd\hat\psi+\cos\hat\theta \dd\hat\phi)\right]^2
\end{aligned}
\end{equation}
Assuming, as in the generic case, $\alpha\equiv - \tilde\phi$, and equating the $g_{3\alpha}$ and $g_{\alpha\alpha}$ components we get
\begin{equation}
	 A_\phi-\delta = - 3 \hat y
\end{equation}
and
\begin{equation}
	\frac 3 2 \tau^2 + 4 \hat y^2 = 2(3-\hat y^2)\quad \then\quad \tau^2 = 4 (1-\hat y^2)
\end{equation}
Assuming $r= r(\hat y)$ and equating the $\dd \hat y^2$ term gives
\begin{equation}
	\frac{\der \ln r}{\der \hat y} = \pm \frac{1}{6(1-\hat y^2)}\quad\then\quad r = \lambda \left(\frac{1-\hat y}{1+\hat y}\right)^{\mp 1/12}
\end{equation}
where $\lambda$ is an arbitrary constant and we fix $\lambda = 1$. We are now able to determine the constant $\delta$ through the equation
\begin{equation}
 	A_\phi = A_t V_\phi + \frac 12 r \der_r \ln T = - \frac 12 \mp 3\hat y
\end{equation}
which fixes the upper choice for the sign and
\begin{equation}
 	\delta = -\frac 12
\end{equation}

In order to bring the metric to the standard $T^{1,1}$ form we set
\begin{equation}
	\hat y = - \cos \tilde\theta 
\end{equation}
which gives
\begin{equation}
	r= \left(\tan \frac{ \tilde\theta} 2 \right)^{1/6} \quad,\quad \tau = 4 \sin^2 \tilde\theta
\end{equation}
and thus
\begin{equation}
	\frac{\dd s_5^2}{L^2} = \frac{1}{6}\big( \dd \tilde \theta^2 + 36 \sin^2 \tilde \theta \dd \tilde \phi^2 \big) + \frac 1 6 \big(\dd\hat \theta^2 + \sin \hat\theta \dd\hat \phi^2 \big) + \frac 1 9 \big( \dd\hat \psi + \cos \hat\theta \dd\hat \phi + 6 \cos \tilde \theta \dd \tilde \phi\big)^2
\end{equation}
which is the $T^{1,1}$ metric up to the trivial rescaling
\begin{equation}
	\tilde \phi \tend \frac 1 6 \tilde \phi
\end{equation}
\section{Asymptotic expansion for half BPS states in  $AdS_5 \times Y^{p,q}$}\label{subcorr}
In this section we study generic asymptotic perturbations of the $AdS_5\times Y^{p,q}$ geometries that preserve
$1/2$ of the bulk supersymmetries. We relax the constraints of \eqref{constraint} which give back $AdS_5\times Y^{p,q}$
and solve the differential equations \eqref{equations} with the boundary conditions that the
solutions approach $AdS_5\times Y^{p,q}$ at large distances
(including also the particular case $c=0$). We will work in the somewhat mixed coordinates 
$(y,\hat y)$ or $(y,\tilde\theta)$ and solve the equation in an expansion for  large $y$, 
with the simplifying assumption  
that the solutions are invariant under shifts in $\tilde \phi$.
We make the following Ansatz for the expansion of our functions,
\begin{align}
& \rho_1 = L \sqrt {\frac 23 (1 - c \hat y)} \left(1 + \rho_1^{(1)}(\hat y)\frac {L^4}{y^2} +\rho_1^{(2)}(\hat y) \frac{L^8}{y^4} \right)\\
&\rho_3 = L \sqrt { \frac{2(2 + a c^2- 6 c \hat y + 3 c^2 \hat y^2)}{9(1-c \hat y)}} \left(1 + \rho_3^{(1)}(\hat y)\frac{L^4}{y^2} + \rho_3^{(2)}(\hat y) \frac{L^8}{y^4} \right)\\
&\tilde\rho = \frac y{\rho_1}\\
&A_t = \frac{1 - a c^2}{2+ a c^2 - 6 c \hat y +3 c^2 \hat y^2} \left(1 + A_t^{(1)}(\hat y)\frac{L^4}{y^2} +A_t^{(2)}(\hat y) \frac {L^8}{y^4} \right)\\
&T = \frac y r \sqrt{\frac{2(a - 3 \hat y^2 +2 c \hat y^3)}{(1-c \hat y)^3}}\left(1 + t^{(1)}(\hat y)\frac{L^4}{y^2} + t^{(2)}(\hat y)\frac {L^8}{y^4} \right)\\
&V_\phi = \frac{ 4 c (1- c\hat y)(a-3 \hat y^2+2 c \hat y^3)}{3(2+ a c^2 -6 c \hat y + 3 c^2 \hat y^2)}\frac{L^4}{y^2}+V_\phi^{(2)}(\hat y)\frac{L^8}{y^4}+ V_\phi^{(3)}(\z)\frac{L^{12}}{y^6}\\
&r= y^{c/2} r_0(\hat y)
\end{align}
This expansion reproduces the $c=0$ limit upon setting $a=3$, as in the previous section.

In these coordinates, the condition \eqref{r0} becomes
\begin{equation}
 	r_0'(\hat y) = \frac{2 + a c^2 - 6 c \hat y +3 c^2 \hat y^2}{4(1- c \hat y)(a-3\hat y^2 + 2 c \hat y^3)}r_0(\hat y)
\end{equation}
The functions $m,n,p$ are given by
\begin{align}
&m = \frac{1}{\rho_3^2[\tilde \rho^2 +(1+A_t)^2\rho_3^2]}\\
&n = \frac{(1+A_t)\rho_1^2}{y^2[\tilde \rho^2 +(1+A_t)^2\rho_3^2]}\\
&p = \frac{\rho_1^2}{y^2[\tilde \rho^2 +(1+A_t)^2\rho_3^2]}\,.
\end{align}
The constraints in \eqref{constraint} and the equations \eqref{ztilde} are satisfied at leading order in $y$.
We rewrite the generic equations \eqref{equations} in polar coordinates dividing them by $T^2$ and exploiting the $U(1)$ symmetry of our solutions
\begin{equation}
\begin{split}
&\frac{y^3}{r^2 T^2} r \der_r (r\der_r n) + \der_y \left(y^3 \der_y n\right) + y^2 \der_y\big[\big(y D n +2y^2m(n-p)\big)\big]\\
&\quad\quad\quad+2 y^2 D (2 n + y  \der_y n + y D n)=0\\
&\frac{y^3}{r^2 T^2} r \der_r (r \der_r m) + \der_y \left(y^3 \der_y m\right) + \der_y \left( y^3 2m D\right)+2 D y^3 (\der_y m + 2  D m)=0\\
&\frac{y^3}{r^2 T^2} r \der_r (r \der_r p) + \der_y \left(y^3 \der_y p\right) + \der_y \big[ y^3  4n y(n-p)\big]+2 D y^3[\der_y p +4n y (n-p)]=0\\
&\der_y \ln T = D\,.
\end{split}
\end{equation}
where
\begin{align}
 	\der_y f (y,\hat y)\equiv \left.\frac{\dd f}{\dd y}\right |_r &= - \frac{c \,r_0(\hat y)}{2 r_0'(\hat y)} \left.\frac{\dd f}{\dd \hat y}\right |_{y}+\left.\frac{\dd f}{\dd y}\right |_{\hat y} \\
	 r \der_r f (y,\hat y) \equiv r \left.\frac{\dd f}{\dd r}\right |_y &=\frac{r_0(\hat y)}{r_0'(\hat y)}\left.\frac{\dd f}{\dd \hat y}\right |_y 
\end{align}

The generic asymptotic solutions to these equation are specified, at each order, by 7 integration constants. As in \cite{Gava:2006pu}, requiring regularity of the solutions implies that not all of them are independent and indeed we have only three independent integration constants. 

For the case of $T^{1,1}$ asymptotics, specified by $c=0$ the first subleading corrections are given by:
\begin{align}
 &\rho_1^{(1)}(\tilde \theta)= - k + C_1 \cos\tilde \theta\\
&\rho_3^{(1)}(\tilde \theta)= \rho_1^{(1)}(\tilde \theta)+k^{(1)}(\tilde \theta)\\
&k^{(1)}(\tilde \theta) = k\\
&A_t^{(1)}(\tilde \theta) = C_2 - 4 C_1 \cos \tilde\theta\\
&t^{(1)}(\tilde \theta)=\frac{L^2\sqrt{2/3}(1 + 9 k)\sin\tilde\theta }{\tan\frac {\tilde\theta}2}\\
&V_\phi^{(2)}(\tilde \theta)= - \frac 8 3 C_2 \sin^2\tilde\theta
\end{align}
while in the generic case we get:
\begin{align*}
 	&\rho_1^{(1)}(\hat y) = \frac{A[2 c^2 K +9  A k +4 \z B C_1 ]}{6(2 + a c^2 - 6 c \z + 3 c^2 \z^2)}\\
&\rho_3^{(1)}(\z)= \rho_1^{(1)}(\z)+k^{(1)}(\z)\\
&k^{(1)}(\z) = \frac{A[4c^2 L K + 9  \big(-2 + 8 c \z - 3 c^3 \z^3 - a c^2(4-c\z)\big)k] }{6(2 + a c^2 - 6 c \z + 3 c^2 \z^2)}\\
&A_t^{(1)}(\z) = \big(\frac{-4c^2A^3 K}{(2 + a c^2 - 6 c \z + 3 c^2 \z^2)^2}+
\frac{2 A}{L}C_2-\frac43 A C_1 +\\&- \frac{9 c^2\ A [a^2 c^2 + \z^2(12 - 26 c\z + 21 c^2\z^2 - 6 c^3\z^3)+ 
        2 a(-2 + 3 c\z - 3 c^2\z^2 + c^3\z^3)]}{2LB} k\\
&t^{(1)}(\z) = \frac{A(4 L -27 k)}{6B}
\end{align*}
where
\begin{align*}
&A = 1 - c\z\\
&B = 2 + ac^2 - 6c\z^2 + 2 c \z^3\\
&K = a - 3 \z^2 + 2 c \z^3\\
&L = 1 - a c^2
\end{align*}
The three arbitrary integration constants, $C_1,C_2, k$ will turn out to be related to the supergravity dual of the flavour and baryon charge of the solutions.
The second order regular solutions are rather complicated.  In general, they will involve new integrations constants together with a inhomogeneous part. The expressions for the inhomogeneous part can be found in the Appendix.

As already noticed, any $c\neq0$ can be reabsorbed by  a redefinition of $\hat y$ and so we set $c=1$.

\section{$U(1)$ charges}\label{charges}
We will now show how the first subleading corrections described in the previous section give rise to the  Kaluza-Klein reduction of type IIB supergravity on the $Y^{p,q}$ manifolds respecting the symmetry of our Ansatz. We will calculate the global charges of the solutions under three $U(1)$ massless KK gauge fields living in $AdS_5$; two of them can be identified with the KK modes of the metric associated to the Killing vectors $\der_{\alpha}$ and $\der_{\hat\psi}$ and which are dual to the flavour and $R$ charges of the dual quiver gauge theory (more precisely to linear combinations of the charges) while the third one is associated to the expansion of the RR 4-form potential on the cohomology of $Y^{p,q}$ and  it is dual to the baryon charge of the gauge theory. Since the third Betti number of such manifolds is one there is only one baryon charge.

In general, the metric on the compact manifold is modified by the metric KK gauge fields as
\begin{equation}
 	\dd s^2 = g_{\alpha \beta} (\dd \xi^\alpha + K^\alpha_I A^I_\mu \dd x^\mu)  (\dd \xi^\beta + K^\beta_I A^I_\mu \dd x^\mu) 
\end{equation}
where $\xi^\alpha$ are coordinates in $Y^{p,q}$ and $x^\mu$ in $AdS_5$ and
\begin{equation}
 	K_I = K^\alpha_I\der_\alpha\qquad I=1,\ldots,n
\end{equation}
are $n$ Killing vectors of $Y^{p,q}$.
In our case, only two gauge fields are turned on and they are associated to $\der_{\alpha}$ and $\der_{\hat\psi}$. We denote the two global gauge charges respectively as $J$ and $Q$. The leading order of the corresponding gauge fields $A_J,A_Q$ is thus given by
\begin{equation}
 	A_J = \frac J {\rhotilde^2}\dd t\qquad A_Q = \frac Q {\rhotilde^2}\dd t\,.
\end{equation}

The metric is modified by the shifts
\begin{align}
 	\dd \hat \psi &\tend \dd \hat \psi + \frac{Q}{\rhotilde^2}\dd t\\
	\dd \alpha &\tend \dd \alpha + \frac{J}{\rhotilde^2}\dd t
\end{align}
to
\begin{multline}
 	\dd s^2 L^{-2} = \frac{1-c\hat y}{6}(\dd\hat\theta^2+\sin^2\hat\theta
      \dd\hat \phi^2)+\frac{1}{w(\hat y)q(\hat y)}
      d\hat y^2+\frac{q(\hat y)}{9}(\dd\hat \psi+\frac{Q}{\rhotilde^2}\dd t+\cos\hat\theta \dd\hat\phi)^2 \\
      + {w(\hat y)}\left[\dd\alpha +\frac J {\rhotilde^2} \dd t-\frac{ac-2\hat y+\hat y^2c}{6(a-\hat y^2)}
         (\dd\hat\psi +\frac{Q}{\rhotilde^2}\dd t+\cos\hat\theta \dd\hat\phi)\right]^2\,.
\end{multline}

Given the expression above for the metric and the solution of the 
equations of motion up to the first sub-leading order we obtain
\begin{align}
 	Q &=  3 C_2-2 C_1,\\
	J &=  \frac 1 2 C_2 -C_1.
\end{align}

Similarly, in the case of $T^{1,1}$ we have

\begin{multline}
	\dd s^2L^{-2} =
\frac{1}{6}\big( \dd \tilde \theta^2 + 36 \sin^2 \tilde \theta (\dd \tilde \phi+\frac{J}{\rhotilde^2}\dd t)^2 \big) + \frac 1 6 \big(\dd\hat \theta^2 + \sin \hat\theta \dd\hat \phi^2 \big)+\\
 + \frac 1 9 \big( \dd\hat \psi+ \frac{Q}{\rhotilde^2}\dd t + \cos \hat\theta \dd\hat \phi + 6 \cos \tilde \theta (\dd \tilde \phi+ \frac{J}{\rhotilde^2}\dd t)\big)^2
\end{multline}
with
\begin{align}
 	Q &=  \frac 3 2 C_2\\
	J &=   -C_1
\end{align}
\subsubsection*{$R$-charge and Reeb vector}
In order to correctly identify the $R$ charge we proceed as in \cite{Berenstein:2002ke,Martelli:2004wu,Herzog:2004tr}. We define the new coordinates
\begin{align}
 	\hat\psi'&=\hat\psi\\
	\beta &= - 6 \alpha+ c \hat\psi
\end{align}
In this coordinate system we can write the metric as a local $U(1)$ fiber over a Kaehler-Einstein manifold and $\hat\psi'$ is a coordinate on the local $U(1)$ fiber.  From
(2.17) of \cite{Martelli:2004wu}, we have
\begin{equation}
\dd\Omega_{Y^{p,q}}^2 = (e^{\hat \theta})^2 + (e^{\hat\phi})^2 + (e^{\z})^2 + (e^\beta)^2
+ (e^{\hat\psi})^2
\end{equation}
where the one forms on $Y^{p,q}$ are,
\begin{equation}
 e^{\hat \theta} = \sqrt{\frac{1-c \z}{6}} d\hat\theta \; , \; \; \;
e^{\hat\phi} = \sqrt{\frac{1-c\z}{6}} \sin\hat\theta \dd \hat\phi \ , 
\end{equation}

\begin{equation}
e^{\z} = \frac{1}{\sqrt{wq}}\dd \z \; , \; \; \; 
e^\beta = \frac{\sqrt{wq}}{6} (\dd\beta + c \cos\hat\theta \dd\hat\phi) \ ,
\end{equation}

\be\label{epsiprime}
e^{\hat \psi'} = \frac{1}{3} (-\dd \hat\psi' - \cos\hat\theta \dd\hat\phi + \z(\dd\beta +c \cos\hat\theta 
\dd\hat\phi)) \ .
\ee

%
As noted in \cite{Berenstein:2002ke}, the $R$-symmetry is identified with a shift in the angular variable 
\be\label{psiR}
\psi_R = -\frac 12 \hat \psi'
\ee
at constant $\beta$. As a consequence, the $U(1)$ $R$-symmetry gauge field is given by
\begin{equation}
 	A_R = -\frac 12 A_Q
\end{equation}
and
\begin{equation}
 	Q_R = -\frac 12 Q\,.
\end{equation}
The associated Killing vector is given by
\begin{equation}
 	K_R = -2 \der_{\hat \psi } - \frac c 3 \der_\alpha
\end{equation}
which coincides with the Reeb vector of the Sasaki-Einstein manifold. Notice that our Reeb vector differs by a factor of 2/3 from the one in \cite{Gauntlett:2004yd},\cite{Martelli:2004wu}.
%
%

\section{The 5-form and baryon charge}\label{5form}
The self-dual Ramond-Ramond field strength $F_{(5)}$ can be written as
\begin{equation}
 	F_{(5)}=\mathcal F_5 + \star_{10} \mathcal F_5\,.
\end{equation}
With our conventions and normalisations, the leading order for $\mathcal F_5$ is given by
\begin{equation}\label{leadingF5}
 	\mathcal F^0_5 = L^4 Vol (Y^{p,q})
\end{equation}
where $Vol (Y^{p,q})$ is the volume form of the unit radius $Y^{p,q}$.
The background metric is perturbed by the KK gauge fields as described in the previous chapter: the field strength is also perturbed in order to satisfy the equations of motion. The corrections are known to be of the form \cite{Berenstein:2002ke,Herzog:2004tr,Benvenuti:2006xg}
\begin{equation}
 	\mathcal F_{5}^{1} = L^4 \dd \left( A_Q \wedge\omega_Q+A_J\wedge\omega_J+A_B\wedge\omega_B\right)\,.
\end{equation}
The $Y^{p,q}$ three forms $\omega_I$ are  defined by
\begin{equation}
 	\dd \omega_I + \iota_{K_I} Vol(Y^{p,q}) = 0
\end{equation}
where $K_I$, $I=J,Q$ is the Killing vector of $Y^{p,q}$ associated with the $A_I$ gauge field. The 3-forms $\omega_{J,Q}$ are clearly defined up to the addition of a closed form.
The 3-form $\omega_B$ is the generator of the one dimensional cohomology of the Sasaki-Einstein manifold and $A_B$ is the gauge field dual to the baryon current of the CFT.
The arbitrary shift by a closed form of the $\omega_{J,K}$ corresponds to the possibility of shifting the mesonic symmetries of the theory by an arbitrary baryonic one.

In the case of generic $Y^{p,q}$ for $c\neq 0$ we obtain the following form for the subleading corrections to $F_{(5)}$,
\begin{multline}
 	\mathcal F_5^1 = - \frac{2}{\rhotilde^3}\dd\rhotilde \wedge\dd t \wedge\bigg\{\frac k 4\left< \left(\sigma^{\hat 3}-3 \hat y (1- \hat y)\dd\alpha \right)\sigma^{\hat 1}\wedge\sigma^{\hat 2}-
\frac 3 {2(1- \z)^2} \sigma^{\hat 3}\wedge\dd\alpha\wedge\dd\hat y\right]+\\
+\frac{Q}{9}\left[ \left(\frac{a - 1 }{3}\sigma^{\hat 3}-\frac{a-2\z(a-1) -3\z^2 +2 \z^3}{2(1-\z)}\dd\alpha \right)\sigma^{\hat 1}\wedge\sigma^{\hat 2}+
\frac {2+ a -6\z+3\z^2} {4(1- \z)^2} \sigma^{\hat 3}\wedge\dd\alpha\wedge\dd\hat y\right]\\+
\frac J 3\left[ \left(\frac{a - 2\z+\z^2 }{3}\sigma^{\hat 3}-\frac{a-2a \z+\z^2}{2(1-\z)}\dd\alpha \right)\sigma^{\hat 1}\wedge\sigma^{\hat 2}-
\frac {a -2\z+\z^2} {2(1- \z)^2} \sigma^{\hat 3}\wedge\dd\alpha\wedge\dd\hat y\right]\bigg\}+\\
+\frac {1} {\rhotilde^2} \dd t\wedge\bigg(-Q \frac {2(1-\z)}{9} \dd \alpha - J \frac {4  (1-\z)}9 \sigma^{\hat 3}\bigg)\wedge\dd\z\wedge\sigma^{\hat 1}\wedge\sigma^{\hat 2}
\end{multline}
while for the case $c=0$ and going to the natural coordinate $(\tilde \theta,\tilde\phi)$ defined by $(\z,\alpha) = (-\cos\tilde\theta,-\tilde\phi)$ we get
\begin{align}
 	\mathcal F_5^1 = - \frac{2}{\rhotilde^3}\dd\rhotilde \wedge\dd t \bigg\{-\frac k 6 &\left[ \left(2 \sigma^{\hat 3}- 6\cos\tilde\theta \dd\tilde\theta\wedge\dd\tilde\phi \right)\sigma^{\hat 1}\wedge\sigma^{\hat 2}-
 3 \sin\tilde\theta\sigma^{\hat 3}\wedge\dd\tilde\theta\wedge\dd\tilde\phi\right]\notag\\
+ \frac{Q}{9}&\left[ \left(\frac 13 \sigma^{\hat 3}- \cos\tilde\theta \dd\tilde\phi \right)\sigma^{\hat 1}\wedge\sigma^{\hat 2}+
\frac 1 {2} \sin\tilde\theta\sigma^{\hat 3}\wedge\dd\tilde\theta\wedge\dd\tilde\phi\right]\notag\\
+\frac{J}{3}&\left[ \left(-\frac 43 \cos\tilde\theta\sigma^{\hat 3}+\frac12(1+7 \cos 2\tilde\theta) \dd\tilde\phi \right)\sigma^{\hat 1}\wedge\sigma^{\hat 2}+\frac12
\sin2\tilde\theta\sigma^{\hat 3}\wedge\dd\tilde\theta\wedge\dd\tilde\phi\right]\bigg\}\notag\\
+\frac {1} {\rhotilde^2} \dd t\wedge&\bigg( Q \frac 2 {9}\dd\tilde\phi+ J \frac 4 {9}  \sigma^{\hat 3}\bigg)\wedge\sin\tilde\theta\dd\tilde\theta\wedge\sigma^{\hat 1}\wedge\sigma^{\hat 2}
\end{align}

%
%
The volume form on $Y^{p,q}$ is given by\footnote{The orientation is chosen to satisfy \eqref{leadingF5}}
\begin{equation}
 	\text{Vol}(Y^{p,q}) = - e^{\z}\wedge e^{\beta}\wedge e^{\hat \theta}\wedge e^{\hat\phi} \wedge e^{\hat \psi'}=\frac{4(1 - c\z) }{9} \dd \hat y \wedge\dd \alpha\wedge \sigma^{\hat 1}\wedge\sigma^{\hat 2}\wedge\sigma^{\hat 3}\,,
\end{equation}
and we define the three forms
\begin{multline}
 	\omega_\pm\equiv e^{\hat \psi'}\wedge(e^{\hat\theta}\wedge e^{\hat\phi}\pm e^{\z}\wedge e^{\beta})=\\=
  \frac 13 \left( 2\sigma^{\hat 3} (1-c \hat  y) - 6 \hat y \dd \alpha\right)\wedge\left(\frac{2(1- c \hat y )}{3}\sigma^{\hat 1}\wedge\sigma^{\hat 2}\mp\frac 13 \dd \hat y\wedge(c \sigma^{\hat 3} + 3 \dd\alpha)\right)
\end{multline}
The local K\"ahler form $J_2$ is given by
\begin{equation}
 	J_2 = e^{\hat\theta}\wedge e^{\hat\phi}- e^{\z}\wedge e^{\beta}=\frac 12 \dd e^{\hat \psi'}
\end{equation}

The closed form $\omega_B$ is given as in \cite{Herzog:2004tr} by
\begin{equation}
 	\omega_B = \frac{9}{8 \pi^2(1-c \z)^2}(p^2-q^2) \omega_-
\end{equation}
With this normalisation and assuming that $A_B = \frac{Q_B}{\rhotilde^2}\dd t$, the baryon charge $Q_B$ is given by
\begin{equation}
	Q_B =  \frac{\pi^2}{2(p^2-q^2)} k\,.
\end{equation}
In the case of $T^{1,1}$ and recalling the change of coordinates $(\z,\alpha) = (-\cos\tilde\theta,-\tilde\phi)$ we get
\begin{equation}
 	\omega_{\pm} = \left(\frac 23 \sigma^{\hat 3} - 2 \cos\tilde\theta\dd\tilde\phi\right)\wedge\left(\frac23 \sigma^{\hat 1}\wedge\sigma^{\hat 2}\pm \sin\tilde\theta\dd \tilde\theta\wedge\dd\tilde\phi\right)
\end{equation}
with
\begin{equation}
 	\omega_B \equiv\frac{9}{8 \pi^2}\omega_-
\end{equation}
and
\begin{equation}
 	Q_B = \frac{2 \pi^2}{3} k\,.
\end{equation}
We now rewrite the expansion of $\mathcal F_5^{1}$ as
\begin{equation}
 	\mathcal F_{5}^{1} = L^4 \dd \left( A_R \wedge\omega_R+A_\beta\wedge\omega_\beta+A_B\wedge\omega_B\right)\,.
\end{equation}
where
\begin{equation}
 	A_R = -\frac 12 A_Q, \qquad\qquad A_\beta = -6 A_J -A_Q
\end{equation}
are the gauge fields associated to the Killing vectors
\begin{equation}
 	K_R = -2 \der_{\hat\psi}-\frac 13 \der_\alpha, \qquad\qquad \der_\beta = -\frac 16 \der_\alpha\,.
\end{equation}

The remaining 3-forms are given by
\begin{align}
 	\omega_R =&-  \frac 16 \omega_+\\
	\omega_\beta =& - \frac{(a-2\z +\z^2)}{18} \sigma^{\hat 3}\wedge \left(\frac 2 3 \sigma^{\hat 1}\wedge\sigma^{\hat 2}-\frac{1}{2(1-\hat y)^2}\dd\alpha\wedge\dd\hat y\right)+\\
&-\frac{a-2a\z+\z^2}{18(1-\z)}\dd \alpha\wedge \sigma^{\hat 1}\wedge\sigma^{\hat 2}\,.
\end{align}

It can be shown without difficulty that they satisfy the expected relations
\begin{align}
 	\dd \omega_R  &+ \iota_{(-2\der_{\hat\psi}-\frac13 \der_\alpha)} Vol(Y^{p,q}) = 0\,,\\
	\dd \omega_J &+ \iota_{\der_\beta} Vol(Y^{p,q}) = 0\,.
\end{align}

\section{Mass in Asymptotically $AdS_5 \times X^5$}
\label{asympmass}
In this Section we derive the expression for the mass in the asymptotically 
$AdS_5 \times Y^{p,q}$ spacetimes under examination. There has been a considerable 
amount of work over the years on the definition of mass and other 
conserved charges in general relativity. The issue becomes even subtler in the
case of the definition of the mass in asymptotically $AdS$ spaces. For example,
the standard expression given in terms of a Komar integral gives a divergent result
in this case, and the procedure of renormalisation is ambiguous. 
We will follow the definition of conserved charges given by
Wald and collaborators \cite{Wald:1999wa,Hollands:2005wt}
which provides a possible general framework for addressing this issue,
and apply it to our case for the computation of the mass. 
Since our solutions mix beyond the leading order $AdS$ and $Y^{p,q}$ coordinates,
it is natural to take a ten dimensional approach for the definition of mass, which has the 
advantage of being relatively simple both from the conceptual and from the technical 
point of view.\\
Another derivation of
conserved charges applicable to spacetime with AdS asymptotics (more precisely asymptotically locally $AdS$
spacetimes) was presented in \cite{Papadimitriou:2005ii}. Using non linear KK mapping one can also uplift this derivation to ten dimensional asymptotically $AdS_5 \times X^5$ backgrounds \cite{Skenderis:2006uy,Skenderis:2007yb}.

The main result of this section is to prove that, with the adopted definition of mass, the expected BPS relation:
\begin{equation}
 	M L = \frac 32 R
\end{equation}
which is a consequence of the $\mathcal N=1$ superconformal algebra
on the field theory side, is satisfied.
\subsection{Definition of charges  in asymptotically $AdS_5 \times X^5$ }
We are dealing with an asymptotically $AdS_5 \times X^5$ spacetime, where $X^5$ is a compact manifold.
It is convenient to choose coordinates such that, defining a radial $AdS$ coordinate $\Omega$, 
$g_{\Omega\Omega}=L^2/\Omega^2$ and $g_{\Omega M}=0$ for $M\neq \Omega$, $M$ denoting a ten dimensional
coordinate. We will also denote the $AdS$ coordinates with $\mu,\nu,\dots$ and the internal coordinates with
$a,b,\dots$. At leading order for large $\Omega$ we have 
\begin{equation}
 	\dd s^2 = \frac {L^2} {\Omega^2} \left[ \dd \Omega^2 - \dd t ^2 + \dd \Omega_3^2 \right] + L^2 \dd s^2 (\Ypq)\,.
\end{equation}
We will keep corrections to orders $\Omega^{2k}$, with $k=0,1,2$ for  the $AdS$ part, $g_{\mu\nu}$,
$k=1,2$ for the internal, $g_{ab}$, and mixed parts, $g_{\mu a}$  respectively. 
There are of course corrections of higher order  in $\Omega$ to the background
5-form given by the volume forms on $AdS_5$ and $Y^{p,q}$ which we will discuss later. 

In general the construction of conserved charges proceeds as follows: 
let us denote for the moment as $\varphi$ the generic field appearing in a Lagrangian $\mathcal L$.
The variation of the Lagrangian with respect to $\varphi$ is given by
\begin{equation}
 	\delta \mathcal L =  E (\varphi) \delta \varphi + \dd \theta (\varphi,\delta \varphi)\,.
\end{equation}
where  $E(\varphi)$ denotes the equations of motion. 
This defines the symplectic potential $\theta$, corresponding to the boundary term that arises from 
integrating by parts in order to remove derivative of $\delta \varphi$. It is a 
9-form in spacetime.

We will be interested in the following asymptotic symmetry generator
\begin{equation}
 	\xi = \frac{\der}{\der t}
\end{equation}
We want to identify the Hamiltonian generator $\mathcal H_\xi$ of such symmetry.
Its value on the desired solution will be our definition of the mass of the metric
\footnote{We are specifying here to a particular symmetry generator since 
we are interested in the mass, but the same procedure con be applied to the most general 
asymptotic symmetry generator \cite{Wald:1999wa}.}
$\mathcal H_\xi$ is defined via its variation with respect to a generic fluctuation
$\delta \phi$, obeying the linearised equations of motion
in a given background obeying the full equations of motion \cite{Wald:1999wa}:
\begin{equation}
 	\delta \mathcal H_\xi =\int_{\der \Sigma} \left ( \delta Q_\xi - \xi \cdot \theta \right)
\end{equation}
where $\Sigma$ is a 9 dimensional submanifold of the 
spacetime without boundary, a ``slice'' corresponding to the vector
field $\xi$. 
By the integral over $\der \Sigma$ we mean a  limiting process 
in which the integral is first taken over the boundary $\der K$ 
of a compact region inside $\Sigma$ and then $K$ approaches $\Sigma$ 
in a suitable manner. 
The 8-form $Q_\xi$ is the Noether charge of the asymptotic symmetry $\xi$. It has a contribution
coming from the gravitational lagrangian:
\begin{equation}
 	Q_{\alpha_1\cdots\alpha_8}^{grav} = - \frac{1}{16 \pi G_{10}} \nabla^{b}\xi^c \epsilon_{bca_1\cdots a_8}\,.
\end{equation}
where $\epsilon=\sqrt{- \det{g}}\,\, \dd^{10} x$ is the volume form.
Also, the gravitational contribution to $\theta$ is:
 \begin{equation}
 	\theta_{a_1\cdots a_9}^{grav} = \frac{1}{16\pi G_{10}}v^a \epsilon_{a a_1 \cdots a_9}
\end{equation} 
with
\begin{equation}
 	v^a = \nabla^b \delta g_b^{\phantom b a} - \nabla^a \delta g_b^{\phantom b b}
\end{equation}

Finally, the RR 5-form contributes both to $Q_\xi$ and $\theta$, giving rise to
a single term in the combination $\delta Q_xi-\xi\cdot \theta$. 
With our normalisation for the 5-form $F_5$, the final result for $\delta {\mathcal H}_xi$
is
\begin{equation}
\delta{\mathcal H}_\xi =\int_{\der\Sigma}\frac{1}{16\pi G_{10}}\left( -\delta Q_\xi^{grav}
-\xi^{a1}\left(v^a \epsilon_{a a_1 \cdots a_9} - 128 F_{a_1\cdots a_5} 
\delta A_{a_6\cdots a_9}\right)\right)
\label{deltamass}
\end{equation}
where $F^{(5)} = \dd A^{(4)}$.

Under mild assumptions \cite{Wald:1999wa}, a necessary and sufficient condition for the existence of ${\mathcal H}_\xi$ 
is the integrability of the equation for $\mathcal H_\xi$ : 
\begin{equation}
 	(\delta_1 \delta_2 - \delta_2 \delta_1 ) \mathcal H_\xi = 0
\end{equation}
\textit{i. e.}
\begin{equation}
 	0 = \xi \cdot \left[\delta_2 \theta (\varphi,\delta_1 \varphi) -\delta_1 \theta (\varphi,\delta_2 \varphi) \right]
\end{equation}
When this condition is satisfied we are guaranteed that an 8-form $I_\xi$ exists whose variation is
\begin{equation}
 	\delta I_\xi=\delta Q_\xi - \xi\cdot \theta 
\end{equation}
The value of the global charge associated with the asymptotic isometry generated by $\xi$ is 
given by a simple ``surface'' integral, up to an arbitrary constant which can be 
determined by fixing the value of the charge for a ``reference solution''.
\begin{equation}
 	\mathcal H_\xi = \int_{\der\Sigma} I_\xi\, + \mathcal H_\xi^0.
\end{equation}
Notice that the 8-manifold $\der\Sigma$ in the present case reduces asymptotically to $S^3\times Y_{pq}$, where $S^3$ is
a 3-sphere of radius $L/\Omega$ inside $AdS_5$. 
The existence of $\mathcal H$ can be explicitly checked for a background with the asymptotic behaviour
we have discussed above for the metric. The expression for $\theta^{grav}$ in our gauge is proportional to  
\begin{equation}
\xi\cdot\theta^{grav}(\delta g)=\left(\Omega^2\delta(g^{tM}\partial_{\Omega}g_{aM}\sqrt g )-g^{MN}\sqrt g
(\partial_{\Omega}\delta g_{MN}-\Gamma_{\Omega M}^P\delta g_{PN})\right)\epsilon_{t\Omega M_1\dots M_8}
\end{equation}
One can verify, using the asymptotic expansion for the metric given before, 
that $\delta_{[1}\theta(\delta_{2]} g)=0$. The crucial fact for this result to hold is
that $g^{MN}\delta g_{MN}={\mathcal O}(\Omega^4)$. This is satisfied
by our BPS solutions, but can be proven to hold more generally, even for non necessarily BPS
solutions of the equations of motion, given an appropriate asymptotic behaviour \cite{future}.
One can similarly verify that the contribution of the 5-form to $\theta$ is integrable.

Once we have verified the existence of $\mathcal H_\xi$, we can define the mass of a generic solution $\mathcal M$ to the equations of motion as the value of $\mathcal H_\xi$ on such a solutions
\begin{equation}
 	M_{\mathcal M} \left.\equiv H_\xi \right|_{\mathcal M}\,.
\end{equation}
\subsection{Calculation of mass and $R$-charge}

We will now proceed to the calculation of the mass and $R$-charge for the solutions we have described in the previous Sections.  We are interested in the dependence of
the mass $M=$ on the integration constants, $C_1$, $C_3$ and $k$. Therefore we will
compute, $\frac{\partial M}{\partial C_i}$ and $\frac{\partial M}{\partial k}$, 
by plugging in (\ref{deltamass}) the expressions for the background given by our solutions.

 
Using the expressions for the leading order, first and second subleading orders for the metric and
the 5-form
given in \ref{charges} and in the Appendix one can calculate
\begin{equation}
\begin{split}
\frac{\partial M}{\partial k}= \int_{S^3\times \Ypq} \left ( \frac{\partial}{\partial k} Q_\xi^{grav} - \xi\cdot \theta^k \right)&=0\\
\frac{\partial M}{\partial C_1}=	\int_{S^3\times \Ypq} \left ( \frac{\partial}{\partial C_1} Q_\xi^{grav} - \xi\cdot \theta^1 \right)&= 2 \frac {\pi L^2}{4 G_5}\\
\frac{\partial M}{\partial C_3}=	\int_{S^3\times \Ypq} \left ( \frac{\partial}{\partial C_2} Q_\xi^{grav} - \xi\cdot \theta^2 \right)&= - 3 \frac {\pi L^2}{4 G_5}
\end{split}
\end{equation}
where $G_5$ is the 5-dimensional Planck constant $G_5=G_{10}/Vol(\Ypq)$ and
\begin{equation}
 	\theta^i_{a_1\cdots a_9}  = \frac{1}{16\pi G_{10}}\left[\left(\nabla^b \der_i g_b^{\phantom b a} - \nabla^a \der_i g_b^{\phantom b b}\right) \epsilon_{a a_1 \cdots a_9} - 128 F_{a_1\cdots a_5} \der_i A_{a_6\cdots a_9}\right]
\end{equation}
with
\begin{equation}
 	\der_i = \frac{\partial}{\partial k}\,,\,\frac{\partial}{\partial C_i}.
\end{equation}
Putting everything together we conclude that:

\begin{equation}
 	M = \frac{\pi L^2}{4G_5} (2C_1-3C_2)=- \frac{\pi L^2 Q}{4 G_5} \,,
\end{equation}
where we have set the integration constant to zero. 
Some comments are in order here. First note that the 8-form to be integrated involves directions
orthogonal to $t$ and $\Omega$. The relevant contribution from the 5-form
is of the type $F^{(5)}_{t \hat y \tilde 1 \tilde 2 \tilde 3}\der_i A^{(4)}_{\phi \hat 1 \hat 2 \hat 3}$, which
turns out to be of order  $\Omega^0$:   $\partial_iA^{(4)}_{\phi \hat 1 \hat 2 \hat 3}$ goes
like $\Omega^{2}$,  and $F^{(5)}_{\Omega\phi \hat 1 \hat 2 \hat 3}=\partial_{\Omega}A^{(4)}_{\phi \hat 1 \hat 2 \hat 3}
\sim \Omega$. 
On the other hand,  $F^{(5)}_{t \hat y \tilde 1 \tilde 2 \tilde 3}$, the dual of the latter,  goes like $\Omega^{-2}$. 
Therefore the 5-form term gives a finite contribution to $\der_i M$.
The gravitational contributions to $\der_i M$ on the other hand contain  terms of order $1/\Omega^2$,
therefore potentially divergent. However the coefficients of these terms turn out to be total
derivatives in the internal coordinates: more precisely, 
the coefficient is proportional to $\frac{\dd}{\dd \hat y} q(\hat y)$, therefore it gives 
vanishing contribution after integrating over  $\hat y$ between the two zeroes of $q(\hat y)$,
$\hat y_1$ and $\hat y_2$.
Again this fact can be proven in more generality than just for our BPS solutions \cite{future}. 

Let us now proceed to verify the BPS relation between the mass $M$ and the R-charge $R$.
With our normalisation of the Reeb vector, the BPS relation is given by
\begin{equation}
 	M L = \frac 3 2 R
\end{equation}
where $R$ is the charge which sources the KK gauge field $A_R$.  The five dimensional equation of motion for its field strength $F^R$ are given by
\begin{equation}
 	\tau_{RR} \, \dd\star_5 F^R = \star_5 J^R\,.
\end{equation}
where $J^R$ is the one-form charge current and $\tau_{RR}$ comes from the KK reduction. Taking the integral of the equation of motion, the total charge $R$ can be read from the flux at infinity of the field strength
\begin{equation}
 	R = \lim_{\rhotilde\tend\infty} \tau_{RR} \int_{S^3(\rhotilde)} F^R
\end{equation}
where $S^3(\rhotilde)$ is the three dimensional sphere in $AdS_5$ at constant $t,\rhotilde$.
In Section \ref{charges} we derived
\begin{equation}
 	A^R \approx - \frac Q {2 \rhotilde^2} \dd t
\end{equation}
at leading order in large $\rhotilde$. 
Following \cite{Barnes:2005bw} we have
\begin{equation}
 	\tau_{RR} = \frac{3}{16\pi G_{10}}\int g_{\psi_R\psi_R}  vol(\Ypq) = \frac{1}{12 \pi G_5} 
\end{equation}
where we have used $g_{\psi_R\psi_R}=\frac 4 9$ as can be seen from \eqref{epsiprime},\eqref{psiR}.
We can now explicitly write the value of the total $R$ charge
\begin{equation}
 	R = -\frac{Q L^3}{12 \pi G_5}Vol(S^3) = \frac 2 3 M L
\end{equation}
which satisfies the expected relation.

Let us mention that we have computed $M$ also using a 5-dimensional definition
involving the intrinsic 5-dimensional Weyl tensor, due to Ashtekar 
and collaborators\cite{Ashtekar:1999jx}
and rederived in \cite{Hollands:2005wt},
\begin{equation}
 	\mathcal H_\xi = M = - \frac{1}{8\pi G_{5}} \int_{S^3}\tilde E_{t t} vol (S^3)
\end{equation}
where 
\begin{equation}
 	\tilde E_{t t} =  \frac 1 2 \Omega^{-2} \tilde C_{ \Omega t t \Omega}\,.
\end{equation}
where $\tilde C_{a b c d}$ is the Weyl tensor of the unphysical metric $\tilde g=\Omega^2 g$.
Beyond leading order $AdS_5$ and $\Ypq$ coordinates mix, so, in general the 
metric on the deformed $AdS_5$ depends on the choice of the 5-dimensional slice inside
the 10-dimensional manifold. The calculation, done using our explicit form of the perturbed
metric and allowing a slice dependence on the internal coordinates,  actually reveals that
the slice dependence drops out in the Weyl tensor and gives
the correct result for the mass, as in the previous 10-dimensional computation. The degree of
generality of this result is under investigation \cite{future}.

\section{Conclusions and open problems}\label{concl}

In this paper we have performed an asymptotic, large distance, analysis of 
1/2 BPS states in IIB supergravity $AdS_5 \times \Ypq$. The corresponding 
differential equations are the same
as those found in \cite{Gava:2006pu}, where 1/8 BPS states  of IIB supergravity on $AdS_5 \times S^5$
were analysed. The difference resides in the boundary conditions,
here we require solutions which
are asymptotic to $AdS_5 \times \Ypq$. They carry non trivial charges
under the asymptotic isometries which are dual to the $R$-charge and one $U(1)$ 
flavour charges of the  quiver gauge theories. We have shown that the charges are consistent with the holographic principle which 
in this case relates $\mathcal N=1$ quiver gauge theories to gravity on $AdS_5 \times \Ypq$.
Of course our analysis was only asymptotic:  we did not address 
the issue of regularity  of the solutions over the full configuration space. One
can analyse the solutions in the opposite regime, near $y=0$, but 
it is difficult to connect this region to the large $y$ region. 
It would be very interesting, although probably quite
hard, due to the complexity of the system of non-linear
partial differential equations governing them, to prove the existence of non-singular
solutions, which then would be   
the exact analog of those found in \cite{bubbling} for the
maximally supersymmetric case.

In the course of the analysis we had to cope with the problem of defining the mass
of the states in the asymptotically  $AdS_5 \times  \Ypq$ spacetime. We adopted 
a ten dimensional approach, which uses the definition of charges given by
Wald and collaborators. It gives a finite (and correct) result. A different ``holographic'' approach to this problem, which uses a detailed
analysis of the KK reduction of the 10 dimensional theory to $AdS_5$ can be found in \cite{Papadimitriou:2005ii, Skenderis:2006uy,Skenderis:2007yb}.
We had indications, however, that, at least  for our backgrounds, 
an expression due to Ashtekar and Das \cite{Ashtekar:1999jx,Hollands:2005wt}, which involves
the intrinsic  Weyl tensor in the deformed $AdS_5$,  also gives the correct result. This brings
about various questions. For example, about the finiteness of Wald et al. expression,
one would like to establish it in general terms, without relying to a particular
form of backgrounds. That is, one would like to prove in general,
assuming just that the equations of motion hold with the asymptotic behaviour
of the fields implied by AdS/CFT correspondence,
that potentially divergent terms are total derivatives in the
internal compact manifold. Similarly, it would be interesting to see under which circumstances
the ten dimensional approach finally coincides with the 5-dimensional one of Ashtekar et al. 
We hope to come back to these issues in a future publication\cite{future}.


\appendix
\section{Second order solutions}
We give here the complete expression for the second order solutions
\begin{multline*}
 \rho_1^{(2)}(\hat y) = -(L^8 (-1 + c \z)^2 ((-4 + 4 a c^2 + 27
k)^2 (-80 + 496 c \z - 584\ c^2 \z^2 - 2696 c^3 \z^3 + 
                    11666 c^4 \z^4\\ - 19494 c^5 \z^5 + 
                    16281 c^6 \z^6 - 6696 c^7 \z^7 + 
                    1080 c^8 \z^8 + 
                    a^3 c^6 (-65 + 72 c \z + 
                          20 c^2 \z^2) + \\+
                    a^2 c^4 (50 + 82 c \z + 159 c^2 \z^2 - 
                          752 c^3 \z^3 + 380 c^4 \z^4) + 
                    a c^2 (-40 - 56 c \z + 756 c^2 \z^2 - 
                          3572 c^3 \z^3 + 6449 c^4 \z^4+\\ - 
                          4536 c^5 \z^5 + 1080 c^6 \z^6))+\\ - 
              8 (-1 + a c^2) (-4 + 4 a c^2 + 
                    27 k) (2 + a c^2 - 6 c \z + 3 c^2 
\z^2)^2 (-20 + c \z (4 - 84 C_1) - 
                    264 c^3 \z^3 (1 + 4 C_1) + \\+
                    120 c^4 \z^4 (1 + 4 C_1) + 
                    c^2 \z^2 (133 + 552 C_1) + 
                    a c^2 (5 - 60 C_1 + 
                          20 c^2 \z^2 (1 + 3 C_1) + 
                          2 c (\z + 54 \z C_1))) +\\+ 
              32 (-1 + a c^2)^2 (2 + a c^2 - 6 c \z + 3 
c^2 \z^2)^2 (-10 + c \z (2 - 84 C_1) + 
                    10 c^4 \z^4 (-1 + 10 C_1 + 27 C_1^2) +\\+ 
                    a c^2 (-2 c \z (17 + 41 C_1) + 
                          10 c^2 (\z + 3 \z C_1)^2 + 
                          5 (3 + 8 C_1 - 6 C_2)) - 
                    2 c^3 \z^3 (1 + 143 C_1 + 270 C_1^2 + 
                          30 C_2) + \\
                    c^2 \z^2 (29 + 252 C_1 + 180 C_1^2 + 
                          90 C_2))))/(4320 (-1 + a 
c^2)^2 (2 + a c^2 - 6 c \z + 3 c^2 \z^2)^3)
\end{multline*}

\begin{equation*}
 \rho_3^{(2)}(\z) = \rho_1^{(2)}(\hat y)+k^{(2)}(\z)
\end{equation*}

\begin{multline*}
 	k^{(2)}(\z)=(L^6 (L - c L \z)^2 (((1 - 
                        c \z) (16 (-1 + a c^2)^2 
(-1 + 
                              c \z) (2 + a c^2 - 6 c \z + 3 c^2
 \z^2)^2 \\(11 + 4 a c^2 - 30 c \z + 15 c^2 \z^2) - 
                        4 (-1 + a c^2) (-4 + 4 a c^2 + 
                              27 k) (2 + a c^2 - 6 c \z + 
                              3 c^2 \z^2)\\ (-44 + 284 c \z - 
                              576 c^2 \z^2 + 426 c^3 \z^3 - 
                              90 c^4 \z^4 - 9 c^5 \z^5 + 
                              a^2 c^4 (-14 + 5 c \z) + 
                              2 a c^2 (-34 + 103 c \z - 
                                    72 c^2 \z^2 + \\+
                                    12 c^3 \z^3)) + (-4 + 4 a 
c^2 + 27 k)^2 (a^3 c^6 (11 + c \z) + 
                              6 a^2 c^4 (-17 + 25 c \z - 
                                    21 c^2 \z^2 + 7 c^3 \z^3) + 
                              3 a c^2 (-36 + 220 c \z +\\- 
                                    324 c^2 \z^2 + 200 c^3 \z^3 - 
                                    51 c^4 \z^4 + 3 c^5 \z^5) + 
                              2 (-22 + 202 c \z - 666 c^2 \z^2 + 
                                    894 c^3 \z^3 - 531 c^4 \z^4 +\\+ 
                                    117 c^5 \z^5))))/((1 - 
                        a c^2) (2 + a c^2 - 6 c \z + 3 c^2 \z
^2)^2) - 
              12 c (4 a^2 c^3 (-4 - 8 C_1 + 
                          7 c (\z + 2 \z C_1) + 6 C_2) + \\+
                    \z (27 k (2 - 8 c \z + 3 c^3 \z^3) (1 + 
                                2 C_1) - 
                          4 c \z (-3 + 2 c \z) (-4 - 
                                8 C_1 + 7 c (\z + 2 \z C_1) + 
                                6 C_2)) + \\+
                    a c (4 c (-7 + 27 k) \z (1 + 
                                2 C_1) + 
                          56 c^4 \z^4 (1 + 2 C_1) - 
                          4 c^3 \z^3 (29 + 58 C_1 - 12 C_2) + 
                          8 (2 + 4 C_1 - 3 C_2)+\\ - 
                          3 c^2 \z^2 (-16 + 9 k - 32 C_1 + 
                                18 k C_1 + 
                                24 C_2)))))/(648 (2 + a c^2 
- 6 c \z + 3 c^2 \z^2)^2)
\end{multline*}
\begin{multline*}
 	t^{(2)}(\z)=(L^8 (-1 + c \z)^2 (2187 k^2 (
-1 + c \z)^2 (-2 + 14 c \z - 9 c^2 \z^2 + 
                    a c^2 (-7 + 4 c \z)) +\\- 
              216 (-1 + a c^2) k (-1 + c \z) (2 + 
                    2 a^2 c^4 + 27 c^2 \z^2 - 45 c^3 \z^3 + \\+
                    18 c^4 \z^4 + a c^2 (-13 + 9 c \z)) + 
              16 (-1 + a c^2)^2 (-9 (-1 + c \z)
^2 (2 + a c^2 - 6 c \z + 3 c^2 \z^2) + \\+
                    8 (-1 + 
                          c \z) (2 + a c^2 - 6 c \z + 3 c^2 
\z^2)^2 C_1 + 1/1 - 
                            a c^2 (3 (-(-1 + a 
c^2)^3 C_2 - 
                              3 (-1 + a c^2)^2 (-1 + c 
\z)^2 C_2 + \\+27 (-1 + c \z)^6 C[
                                  3] - (-1 + 
                                    a c^2) (-1 + c \z)^3 (
-4 + 4 a c^2 + (9 - 9 c \z) C_2))))))/(216 (1 - 
              a c^2)\\ (2 + a c^2 - 6 c \z + 3 c^2 \z^2)^3)
\end{multline*}

\providecommand{\href}[2]{#2}\begingroup\raggedright\endgroup
\end{document}